\documentclass[11pt,a4paper]{iopart}
\usepackage[english]{babel}

\usepackage{iopams} 

\expandafter\let\csname equation*\endcsname\relax
\expandafter\let\csname endequation*\endcsname\relax
\usepackage{amsmath}
\usepackage{amssymb}
\usepackage{color}
\usepackage{subcaption}
\usepackage{ifthen} 
\usepackage{ifpdf}
\ifpdf
	\usepackage[pdftex]{graphicx}
	\usepackage[pdftex]{hyperref} 
	\hypersetup{colorlinks=true,
		linkcolor=blue,
		citecolor=blue,
		urlcolor=blue,
	  pdfauthor={LIER ANDREJ andrej.lier@ipp.mpg.de},
	  pdfsubject={2022 A. Lier Mitigated alpha runaway paper},
	  pdftitle={2022 A. Lier Mitigated alpha runaway paper}
	}
	\DeclareGraphicsExtensions{.pdf}
\else
	\usepackage[dvips]{graphicx}
	\usepackage{epsfig}
	\DeclareGraphicsExtensions{.eps}
	\usepackage{url}
\fi
\graphicspath{{figures/}}

\usepackage[numbers,sort&compress]{natbib} 
\usepackage{hypernat}

\usepackage[all]{hypcap}
\usepackage[titletoc,title]{appendix}

\newcommand{\unit}[1]{\ensuremath{\,\mathrm{#1}}}

\newcommand{\qmark}[1]{{``#1''}}

\newcommand{\labelemph}[1]{``#1''}

\begin{document}
\title[The impact of alpha particles on runaway electron dynamics in ITER]{The impact of fusion-born alpha particles on runaway electron dynamics in ITER disruptions}

\author{A.~Lier$^1$, G.~Papp$^1$, Ph.~W.~Lauber$^1$, I.~Pusztai$^2$, K.~S\"arkim\"aki$^1$ and O.~Embreus$^2$}

\address{
$^1$Max Planck Institute for Plasma Physics, D-85748 Garching, Germany\\
$^2$Department of Physics, Chalmers University of Technology, SE-41296 Gothenburg, Sweden\\
}
\ead{\href{mailto:andrej.lier@ipp.mpg.de}{andrej.lier@ipp.mpg.de}}

\begin{abstract}
In the event of a tokamak disruption in a D-T plasma, fusion-born alpha particles take several milliseconds longer to thermalise than the background. As the damping rates drop drastically following the several orders of magnitudes drop of temperature, Toroidal Alfv\'en Eigenmodes (TAEs) can be driven by alpha particles in the collapsing plasma before the onset of the current quench. We employ kinetic simulations of the alpha particle distribution and show that the TAEs can reach sufficiently strong saturation amplitudes to cause significant core runaway electron transport in unmitigated ITER disruptions. As the eigenmodes do not extend to the plasma edge, this effect leads to an increase of the runaway electron plateau current. Mitigation via massive material injection however changes the Alfv\'en frequency and can lead to mode suppression. A combination of the TAE-caused core runaway electron transport with other perturbation sources could lead to a drop of runaway current in unmitigated disruptions.
\end{abstract}

\section{Introduction}
A concern for ITER operation is the threat of the generation of a runaway electron (RE) beam following a plasma disruption~\cite{hollmann2015status,breizman2019physics,boozer2018pivotal}. Mitigation systems~\cite{lehnen2015disruptions} at current development status are predicted~\cite{svensson20magper,vallhagen2022spi} to not be able to confidently mitigate a RE beam generated from a disrupting nuclear phase ITER plasma, representing severe risks to the device integrity~\cite{reux2015runaway,matthews2016melt}. In this paper we discuss an inherent mechanism, which could aid disruption mitigation efforts, but my also aggravate the issue. The phenomenon was originally introduced and investigated in a previous proof-of-principle study~\cite{lier21alpha} and is followed up in this paper.

A plasma disruption~\cite{boozer12disruptions} is an abrupt and uncontrolled discharge termination, which eventually results in a release of the energy stored. With the plasma temperature $T$ dropping many orders of magnitude within milliseconds, the plasma resistivity rises rapidly: $\sigma \propto T^{-3/2}$~\cite{wesson04tokamaks}. The current however cannot vanish from the toroidal plasma on the same time scale as the temperature drops, which leads to the induction of a toroidal electric field. If the electric field grows above a threshold (the $E_\text{c}$ critical electric field~\cite{hesslow18ec}), it can accelerate part of the electron population towards relativistic energies and thus convert a significant fraction of the initial, pre-disruption plasma current into runaway current~\cite{boozer2015theory,breizman2019physics}. For high-current tokamaks like ITER~\cite{shimada07chapter1}, with a maximum plasma current $\approx 15\unit{MA}$, this could generate a RE beam that is able to melt plasma facing components~\cite{matthews2016melt}, or potentially cause sub-surface damage to cooling pipes. 

Traditional mitigation approaches like Massive Material Injection (MMI)~\cite{reux2015runaway,nardon17mgi,baylor15spi,baylor16spi,reux22spi} raise the fraction of plasma energy radiated away in isotropic fashion and elevate the threshold for RE generation. MMI in itself however may not be sufficient to solve the RE problem in reactor-scale tokamaks~\cite{svensson20magper,vallhagen2022spi}. For this reason additional systems, like Resonant Magnetic Perturbations~\cite{lehnen08res,finken07res,papp15magper} (RMPs) or Passive Helical Coils~\cite{smith13passive,weisberg21passive,tinguely21sparc,izzo22phc} (PHCs) are also pursued. Both concepts rely on externally perturbing the confining magnetic field structure and induce a radial transport of runaways. Sufficiently strong perturbations can hinder the formation of a beam and have been investigated in theory~\cite{fulop06destabilisation,pokol2008quasi,fulop09magnetic,komar2012interaction,komar13electromagnetic,pokol14quasi,aleynikov2015stability,liu2018role} and observed experimentally~\cite{yoshino99res,gill02res,zeng13experimental,papp14interaction,liu2018effects,lvovskiy2018role,heidbrink2018low,spong18first,lvovskiy2019observation,ficker17res}. The application of RMPs on current day tokamaks shows mixed results in causing RE losses~\cite{lehnen08res,finken07res,riccardo18jet}, with a major obstacle being the limited penetration depth and insufficient reach towards the predominantly core-generated REs~\cite{papp11runaway,papp12rmp,papp15magper,svensson20magper}. In this work we will investigate a passive and inherent mechanism, which generates core-localized perturbations enhancing core RE transport, and could assist the above mentioned mitigation attempts.

The fusion-born alpha particles in the D-T operation phase of ITER represent an energetic particle (EP) species, with free energy available to drive marginally unstable modes in the quiescent phase~\cite{pinches15ions,schneller16alfven}. The destabilizing effect is counteracted by Landau damping effects coming from the bulk plasma. In the initial phase of a disruption - the thermal quench (TQ) - the temperature-sensitive ion Landau damping~\cite{fu89excitation} strongly decreases and was shown~\cite{lier21alpha} to allow the resonant alpha particle drive to destabilize Toroidicity-induced Alfv\'en Eigenmodes~\cite{fu89excitation} (TAEs) in the weakly damped plasma. The perturbations reach amplitudes significant for RE suppression~\cite{helander00suppression,papp15magper,svensson20magper} and occur during the formation of the RE beam~\cite{lier21alpha}.

We expand the preceding study by including the effects of MMI disruption mitigation systems, a model for post-disruption alpha particle transport, and a self-consistent calculation of the plasma disruption including runaways, addressing the TAE impact on the RE generation and the runaway plateau current. Beginning with a treatment of the alpha particles in section~\ref{sec:coll_alpha}, we introduce an analytical alpha model distribution valid for the thermal quench. In section~\ref{sec:equil} we calculate the plasma equilibrium, the Alfv\'enic mode spectrum and the damping in the post-disruption plasma. The interaction between alphas and TAEs is the subject of section~\ref{sec:wp_int}. The results are then used in section~\ref{sec:dream} for a self-consistent calculation of the ITER disruption, and the results are discussed in section~\ref{sec:discussion}.

\section{Collisional alpha particle dynamics in an ITER plasma thermal quench \label{sec:coll_alpha}} 
ITER is planned to eventually operate with a deuterium-tritium (D-T) plasma, producing alpha particles with a birth kinetic energy of 3.5~MeV. Through collisional thermalization with the background plasma these energetic alpha particles help sustain the plasma temperature, but once thermal, they need to be removed from the system to avoid diluting the fuel~\cite{reiter91heash}. The balance of birth and thermalization creates a steady state energy distribution, for which detailed theoretical predictions for ITER exist~\cite{pinches15ions,moseev19alphas}. In the case of a plasma disruption however, this steady state is forcefully changed as the fusion process stops and plasma parameters suddenly change. In the following we present a model describing the evolution of the alpha particle distribution in a thermal quench.

\subsection{Reduced energetic tail model \label{sec:model}} 
Alpha particles from the D-T fusion emerge isotropically at a super-thermal birth velocity $v_{\alpha 0}$. In a uniform plasma the local alpha velocity space distribution $f$ can be modeled with the Fokker-Planck equation~\cite{helander2005collisional,gaffey76energetic} written in the form valid for isotropic, suprathermal ion species:
\begin{equation}
\frac{\partial f}{\partial t} = \frac{1}{\tau_s} \frac{1}{v^2} \frac{\partial \left(v^3 + v_c^3\right) f}{\partial v} + \frac{S}{v^2} \delta(v-v_{\alpha 0}),
\label{eq:kin}
\end{equation}
where the velocity distribution $f$ depends on time $t$ and velocity $v$, as prescribed by the collisional slowing-down time $\tau_s$, the cross-over velocity $v_c$~\cite{gaffey76energetic}, the source strength $S$ and the Dirac-function $\delta(v-v_{\alpha 0})$. $v_c$ and $\tau_s$ are defined as
\begin{align}
v_c^3 &= v_{te}^3 \frac{3 \sqrt{\pi} m_e}{4 m_\alpha} Z_1 ~ \propto ~ \frac{T_e^{3/2}}{n_e} \sum_i Z_i^2 n_i   , \\
\tau_s &= \frac{3 m_\alpha m_e}{16 \sqrt{\pi} (Z_\alpha)^2 e^4 \mathrm{ln}\Lambda}\frac{v_{te}^3}{n_e} ~ \propto   ~ \frac{T_e^{3/2}}{n_e},\label{eq:tau}\\
Z_1 &=\sum_i \frac{n_i Z_i^2 m_\alpha}{n_e m_i}, \label{eq:Z1}
\end{align}
where $m_\alpha$ is the alpha particle mass, $T_e$ the electron temperature, $m_e$ is the electron mass, $Z_\alpha = 2$ is the alpha particle charge number, $e$ is the Coulomb charge, $\ln\Lambda$ the Coulomb logarithm and $Z_1$ is the effective charge number. Note that there is also a weak dependence on temperature and density in the Coulomb logarithm that is not displayed in the proportionality.

Equation~(\ref{eq:kin}) is valid when small-angle Coulomb collisions dominate and macroscopic accelerating forces are lacking. The alpha species is assumed to be a minority (i.e. the effect of self-collisions is neglected) and the background to be thermal, fulfilling $v_{ti} \ll v_{\alpha 0} \ll v_{te}$, where $v_\text{t,\{i,e\}}=\sqrt{2T_\text{\{i,e\}}/m_\text{\{i,e\}}}$ is the thermal velocity for electrons $e$ and ions $i$ with temperature $T$ and mass $m$. The slowing-down time is evaluated as the inverse of the alpha-electron collision time, while the cross-over velocity is obtained via alpha-ion collisions, rearranged to formally represent a threshold velocity above which the electron drag dominates the ion drag. The known steady state solution to eq.~(\ref{eq:kin}) is the slowing-down distribution~\cite{gaffey76energetic}:
\begin{equation}
f_0 (v) = \frac{\tau_{s0} S_0}{v^3 + v_{c0}^3} \mathrm{U} (v_{\alpha 0}-v),
\label{eq:fsd}
\end{equation}
where $\rm U$ is the unit step function limiting the end of the distribution to the birth velocity. In our modeling, this distribution will be the initial state prior to the thermal quench initiating, hence referred to with an additional subscript 0, meaning $t=0$.

A plasma thermal quench causes the temperature to drop from tens of keVs to the eV level on a millisecond timescale. Meanwhile, influx and consecutive ionization of material can increase the $n$ charged particle number density  of the plasma, where $n = n_e + \sum_i n_i Z_i$, with the electron density $n_e$, the ion density $n_i$ and the ion charge number $Z_i$ of species $i$. Due to quasineutrality, $n=2n_\text{e}$ applies. These main plasma parameter changes explicitly affect the alpha particle dynamics (equation~(\ref{eq:kin})) through the parameters $v_c$ and $\tau_s$.

At $t=0$ the thermal quench initiates, eliminating the fusion source $S_0$ for $t>0$. Eq.~(\ref{eq:kin}) can now be solved by some generic function $F$
\begin{gather}
f(v,t)  = \frac{1}{v^3+v_c^3} F\left( \int^t \frac{\mathrm{d}t}{\tau_s} + \int^v \frac{v^2 \mathrm{d} v}{v^3+v_c^3} \right) = \label{eq:vc} \notag \\ 
= \frac{1}{v^3+v_c^3}  F  \left( \underbrace{\int^t \frac{\mathrm{d}t}{\tau_s} + \frac{1}{3}\mathrm{ln}(v^3+v_c^3)}_{\textstyle \equiv G(v,t)} \right).
\label{eq:1}
\end{gather}
Employing the initial condition, we require $f(t=0) = f_0$, which also allows us to drop the time-dependency in $G(v,t)$ and express $v$ in terms of $G_0 \equiv G(v,t=0)$,
\begin{equation*}
v = (\mathrm{e}^{3G_0} - v_c^3)^{1/3}.
\end{equation*}
Through rearrangement of the initial condition, we obtain a general time-dependent solution
\begin{align}
F(G_0) = \tau_{s0} S_0 \frac{ v^3+v_c^3}{v^3 + v_{c0}^3 } \mathrm{U} (v_{\alpha 0}-v) =  \frac{\tau_{s0} S_0}{1+(v_{c0}^3 -v_c^3)e^{-3G_0}} \mathrm{U}  \left(v_{\alpha 0}  -v \right).
\label{eq:2}
\end{align}
Restoring the time-integral in $G(v,t)$, one arrives at
\begin{align}
f(v,t) & = \frac{1}{v^3 + v_c^3} \frac{\tau_{s0} S_0}{1+(v_{c0}^3 -v_c^3)e^{-3G(t)}} \mathrm{U} (v_\alpha-v) = \notag \\
& = \frac{\tau_{s0} S_0}{v^3 + v_c^3 ( 1- e^{-3 \int \tau_s^{-1} \mathrm{d}t } ) + v_{c0}^3 e^{-3 \int \tau_s^{-1} \mathrm{d}t }} \mathrm{U} (v_\alpha-v).
\label{eq:3}
\end{align}
Note that since the fusion source is disabled, the new cut-off velocity $v_\alpha$ is now time-dependent and given by
\begin{equation}
v_\alpha = \left[ (v_{\alpha 0}^3 + v_c^3)e^{-3 \int_0^t \tau_s^{-1} \mathrm{d}t} - v_c^3 \right]^{1/3}, 
\label{eq:v_cut}
\end{equation}
through a rearrangement of $G(v_{\alpha 0},t)$. At the initial time-point $v_{c} = v_{c0}$ and $v_\alpha = v_{\alpha 0}$ holds, hence we obtain $f(v,0) = f_0$. For our purposes we further approximate the step function with the aid of the complementary Error function Erfc$(x)$ and account for a velocity spread $\Delta v$ at birth~\cite{brysk73fusion}:
\begin{equation}
\mathrm{U} (v_\alpha-v) \longrightarrow \frac{1}{2}\mathrm{Erfc} \left( \frac{v - v_\alpha}{\Delta v}\right).
\label{eq:erfc}
\end{equation}
The velocity spread $\Delta v$ in eq.~(\ref{eq:erfc}) is calculated as an alpha particle thermal velocity corresponding to the background temperature $\Delta v = \sqrt{2 T/m_\alpha}$. 

In a thermal quench, \[ \lim_{t\to\infty} \exp\left\{-3 \int_0^t \tau_s^{-1}(t) \mathrm{d}t\right\} = 0 \] and \[ \lim_{t\to\infty} v_c^3(t) \approx 0, \] but because the exponential approaches zero faster, the inside of the bracket $v_\alpha = [\cdot]^{1/3}$ becomes negative and $v_\alpha$ imaginary. As this is unphysical, the model is only valid until $v_\alpha$ reaches zero, which means, that the energetic tail has ceased to exist. This occurs, because there is no model for the thermal Maxwellian included in the derivation above.  

Important moments of the velocity distribution yield the alpha density $n_\alpha$ and the alpha pressure $p_\alpha$:
\begin{gather}
n_\alpha = \int \mathrm{d}v ~ f (v,t) ,\label{eq:n}  \\ 
p_\alpha = \frac{m_\alpha}{3} \int \mathrm{d}v ~ v^2 f (v,t). \label{eq:p}
\end{gather}

Equations (\ref{eq:3})-(\ref{eq:erfc}) constitute a model, prescribing the velocity space evolution of an ensemble of fusion alphas under changing plasma parameters.
The model relies on the assumption and conservation of isotropy: both the alpha birth and slowing-down process can be considered isotropic to a good degree~\cite{fasoli07chapter}. A disrupting plasma will also generate a directional electric field that could eventually break the validity of this model. In the following we will validate against a numerical solution and discuss the induced electric field. 

\subsection{Validation of the reduced energetic tail model \label{sec:valid}} 
CODION~\cite{embreus15numerical} is a numerical Fokker-Planck solver able to calculate an ion distribution under the influence of an external electric field and small-angle collisions with thermal background populations. It has been equipped with a fusion alpha source~\cite{lier21alpha} and is therefore a good fit to validate our model and test the isotropy assumption. With its ability to resolve the ion distribution in pitch-angle, the influence of electric fields can be studied.

We mimic the initial conditions of an undiluted, 15~MA ITER D-T plasma in the core with an electron temperature of $T_{e0} = 25$~keV, an ion temperature of $T_{i0} = 21$~keV, an electron density of $n_{e0} = 10^{20}$~m$^{-3}$ and a 1:1 D-T ion composition $n_{D0} + n_{T0} = n_{e0}$. The reaction rate for D-T fusion~\cite{NRL}
\begin{equation}
\langle\sigma v\rangle = 3.68 {\cdot} 10^{-18}\, T_{i0}^{-2/3} \mathrm{exp} \left\{ -19.94~T_{i0}^{-1/3} \right\}~\mathrm{m}^{-3} \mathrm{s}^{-1},
\end{equation}
(where $T_{i0}$ is given in keV units) can be used to estimate the alpha particle source magnitude
\begin{equation}
S_0 = n_{D0} n_{T0} \langle\sigma v \rangle.
\label{eq:source}
\end{equation}
The alpha density is calculated to be $n_{\alpha} \approx 0.9 \cdot 10^{-2} n_{e0}$ and matches predictions~\cite{polevoi02iter}. Its diluting effect on the ion density is neglected in evaluating the $S_0$. Since the source is turned off for $t>0$ and no sink is employed, $n_\alpha$ is conserved by our model as well as in CODION. 

The thermal quench is described with an exponential decay in temperature~\cite{wesson04tokamaks,smith08hot}:
\begin{equation}
T_1 = T_f + \left[T_{e0} - T_f \right] \mathrm{e}^{-t_N},
\label{eq:Texp}
\end{equation}
with a final temperature $T_f = 10\unit{eV}$, equal ion and electron temperatures $T_1$ and where we have introduced $t_N \equiv t/t_{\rm TQ}$ with the exponential decay time $t_{\rm TQ}$. The normalized time $t_N$ is a useful metric as it represents a temperature (since, for a given $T_{e0}$ and $T_f$, it corresponds to a temperature -- see eq.~(\ref{eq:Texp})) and is independent of $t_{\rm TQ}$. We model the influx of material coming from mitigation systems with a step-function increase in density, representing the injection via the post-disruption electron density $n_{e1}$: 
\begin{equation}
n_{e0} \leq n_{e1} = n_{e0} + n_\text{D1} + n_\text{Ne1},
\label{eq:ne1}
\end{equation}
where $n_\text{D1}$ and $n_\text{Ne1}$ is the added density of deuterium and neon respectively. Instead of establishing the complicated temperature evolution of the injected deuterium and neon, we assume one temperature for simplicity and avoid high dimensionality scans. The exact temperature of the impurities is not important from the alpha-drive point of view (especially that temperature evolution is predesrcibed) and is unlikely to play a role in the damping either. The injected material is therefore modeled as a Maxwellian fixed at 10~eV and as a singly ionized species.

We note that this approach is a simplification of the MMI dynamics. In the following we will show that the most interesting part of the dynamics happens after the onset of the thermal quench, which requires the MMI material to have already spread well enough to trigger the thermal collapse. While numerical tools exist which model the details of MMI injection and transport (such as nonlinear MHD codes), as the ITER MMI strategy is still under active development~\cite{jachmich22iter}, modeling this would require a multidimensional parameter scan with expensive numerical codes. With the simplified model we retain the main important aspects of the MMI: the triggering of the TQ, and the modification of the Alfv\'en speed and damping.

\begin{figure}
\includegraphics[width = 0.53\linewidth]{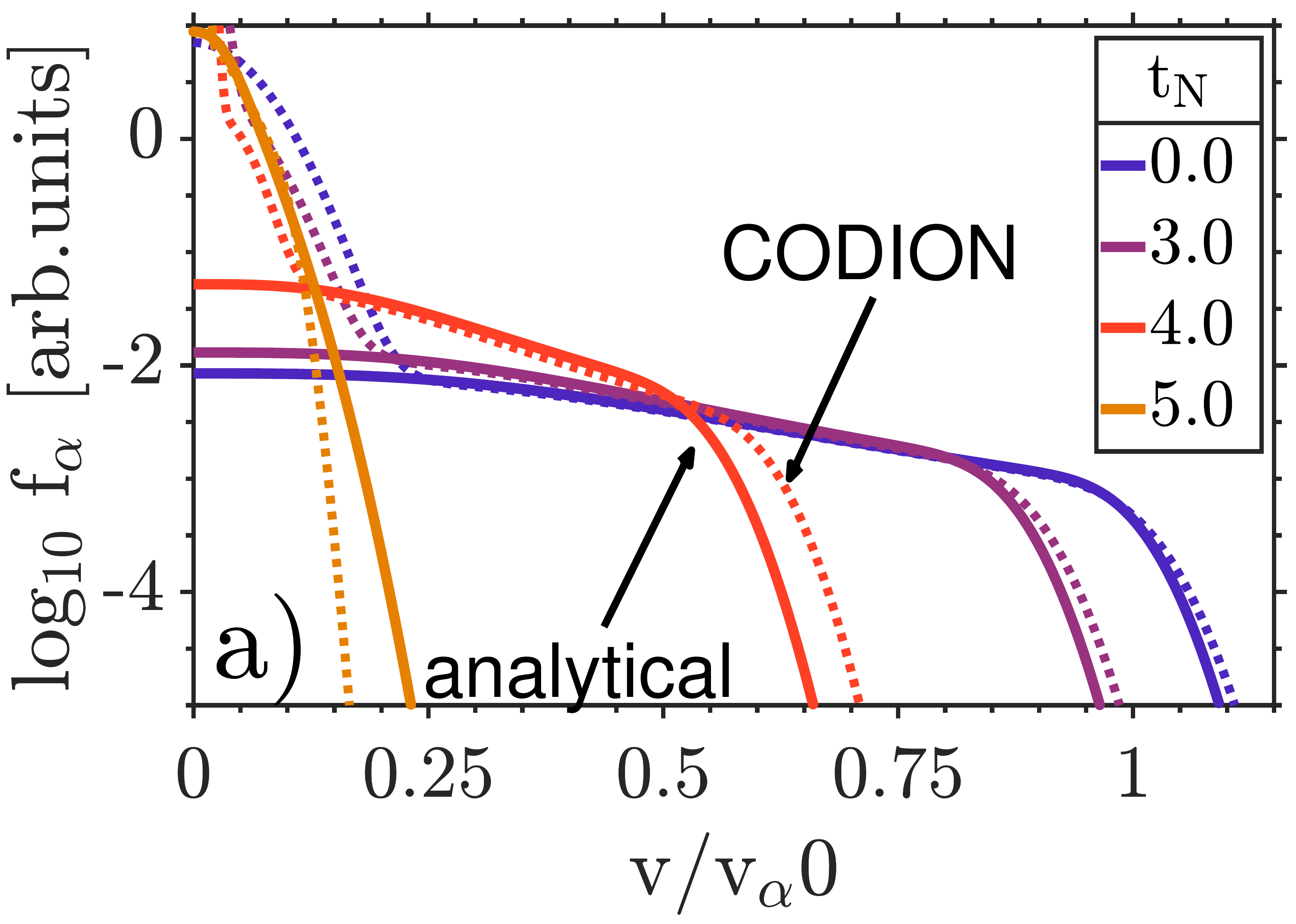}\hfill
\includegraphics[width = 0.45\linewidth]{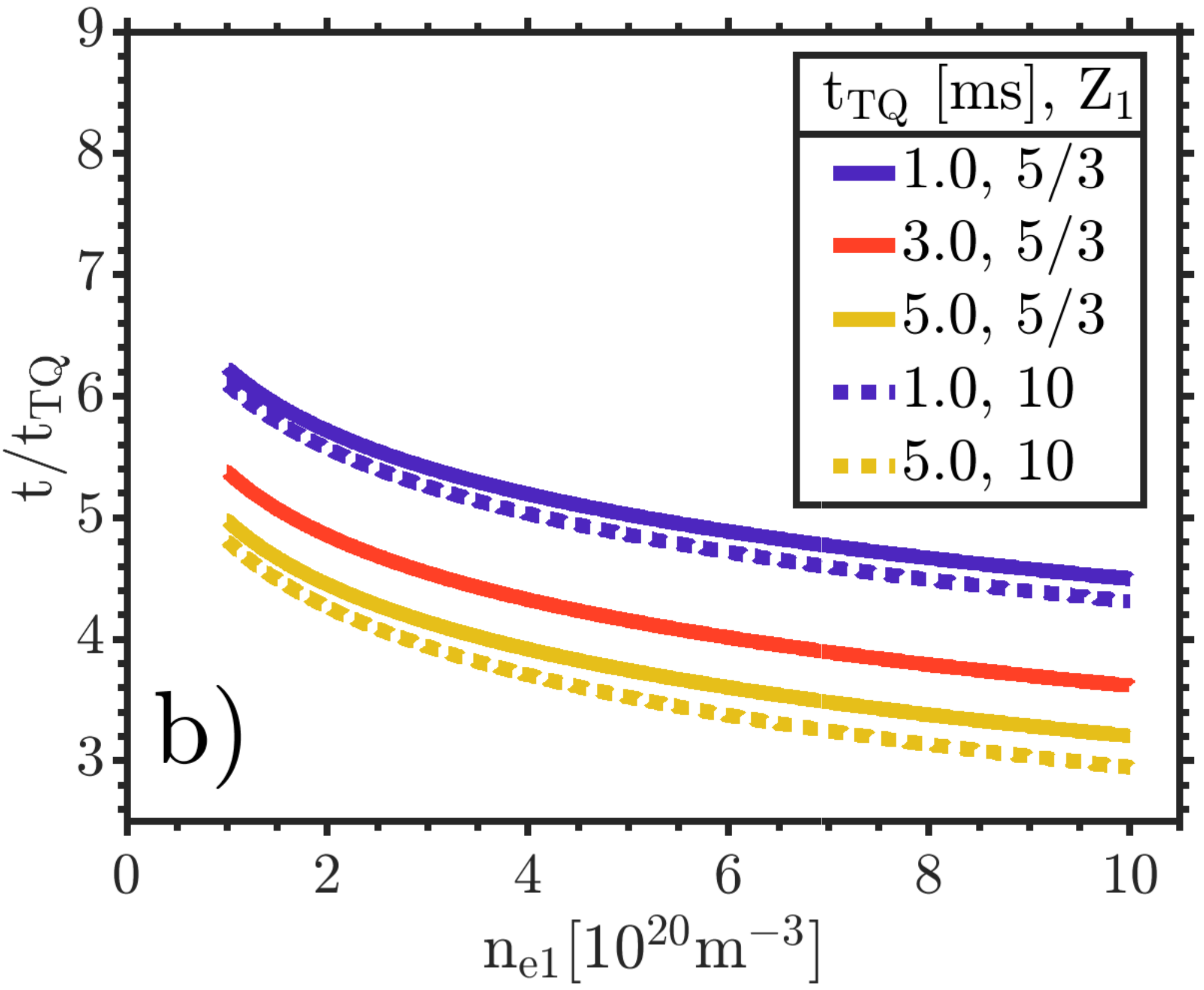}
\caption{(a) The isotropic alpha particle velocity distribution of an exponentially cooling ITER core plasma as computed by CODION (dashed) and the analytical model (solid). Color indicates time. The x-axis is normalized to the birth velocity $v_{\alpha 0}$. As the exponential thermal decay with $t_{\rm TQ} = 1$~ms is initiated, the electron density is increased instantly to $n_{e1} = 4 n_{e0}$ by a singly ionized 1:1 neon+deuterium injection. Note that both solutions conserve particle density, but only CODION includes the Maxwellian bulk of the distribution. (b) Figure shows the time-point (normalized to $t_\mathrm{TQ}$) when the energetic tail has thermalized with the plasma background as a function of post-disruption electron density $n_{e1}$ and the plasma composition represented by $Z_1$ (for details, see text). Color indicates different $t_\mathrm{TQ}$ thermal quench times.}
\label{fig:dis_val}
\end{figure}

For the validation case we inject $n_\text{Ne1} = n_\text{D1} = 3/2 n_{e0}$, elevating to an electron density of $n_{e1} = 4 n_{e0}$ and use the thermal quench time of $t_{\rm TQ} = 1$~ms. The analytical results (previous section) are compared to the CODION simulation in figure~\ref{fig:dis_val}a). Note that CODION includes the Maxwellian bulk of the distribution (while the analytical tail model doesn't), but otherwise we observe a good agreement between the analytical and numerical solutions. Initially, $v_{c0}/v_{\alpha} \approx 0.44$ holds and the most energetic alphas (ones with $v>v_{c0}$) mainly collide with electrons.
The low collisionality of energetic particles allows the energetic tail to withstand the deceleration for approximately $2 t_N$, before experiencing an accelerated cooling as the background temperature decreases further. This acceleration is represented by the time-integral in eq.~(\ref{eq:v_cut}) for $v_\alpha$, with the governing collision time scale decreasing as $\tau_s \propto T_1^{3/2}$. As no particles are born at $v_\alpha$ anymore and the density is conserved, the deceleration begins piling up the particles at lower energies. Due to the high energy that the alpha particles are born at, the EP distribution for $t_N = 5$ is similar in shape to the initial Maxwellian of 25~keV (see figure~\ref{fig:dis_val}a), while the background temperature for this time-point has reached roughly 400~eV already.

A parameter space $[n_{e1}, Z_1, t_{\rm TQ}]$ is set up in order to investigate the thermalization of the alpha particles as a function of thermal quench scenario. $Z_1$ (eq.~(\ref{eq:Z1})) is an effective ion charge weighted by mass and used to represent the plasma composition, with a pure D-T plasma having $Z_1 = 1.66$ and the validation case (above) yielding $Z_1 = 2.6$. From equation~(\ref{eq:v_cut}) we calculate the time-point of $v_\alpha$ reaching zero and display the results of this calculation in the parameter space in figure~\ref{fig:dis_val}b. With an increase in electron density the energetic tail slows down \qmark{quicker} in reference to the background temperature ($t_N \sim T_1$) and can be explained by a reduction of the slowing-down time $\tau_s$. Different quench times yield different results as there is a growing deviation of elapsed time $t=t_N t_\mathrm{TQ}$ to $\tau_s$ (independent of $t_\mathrm{TQ}$). Essentially, a slower quench leaves the alphas more time to thermalize, before a certain bulk temperature is reached ($\sim t_N$). We also observe that raising $Z_1$ accelerates the slowing-down process marginally. This is caused by the increasing cross-over velocity ($v_c \propto Z_1^{1/3}$) and thus increasing the velocity space fraction in which alpha-ion collisions dominate over alpha-electron collisions. The effect however remains minor due to the weak dependency.

The analytical model is stated for a pitch-independent distribution, whose isotropy can be broken by a directional electric field. For the 15~MA ITER case, a pre-disruption electric field of the order of $E_\mathrm{pre} \approx 0.01\unit{V/m}$ is present, but neglected due to its low magnitude. When a hot plasma cools down, its resistivity $\sigma \propto T_1^{-3/2}$ rises, causing the induction of a strong and directional electric field. The current quench typically occurs on a timescale about an order of magnitude longer than the thermal quench. We will now show, that the delay between the temperature drop and electric field rise is long enough to justify the alpha particle isotropy assumption for the entire thermalization process (in the relevant parameter regimes studied in this paper). 

\begin{figure}
\includegraphics[width = 0.95\linewidth]{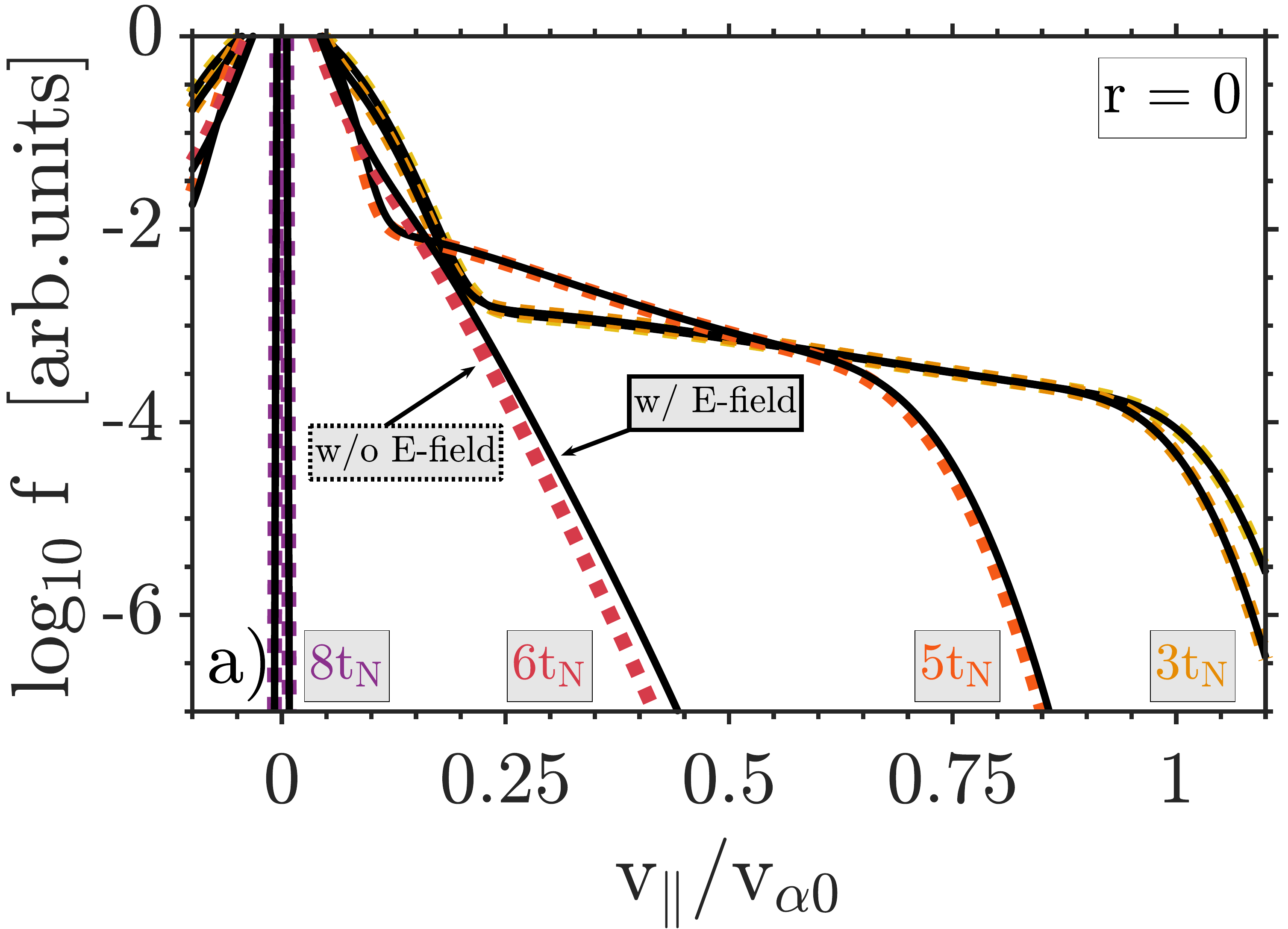}  \vfill
\includegraphics[width = 0.48\linewidth]{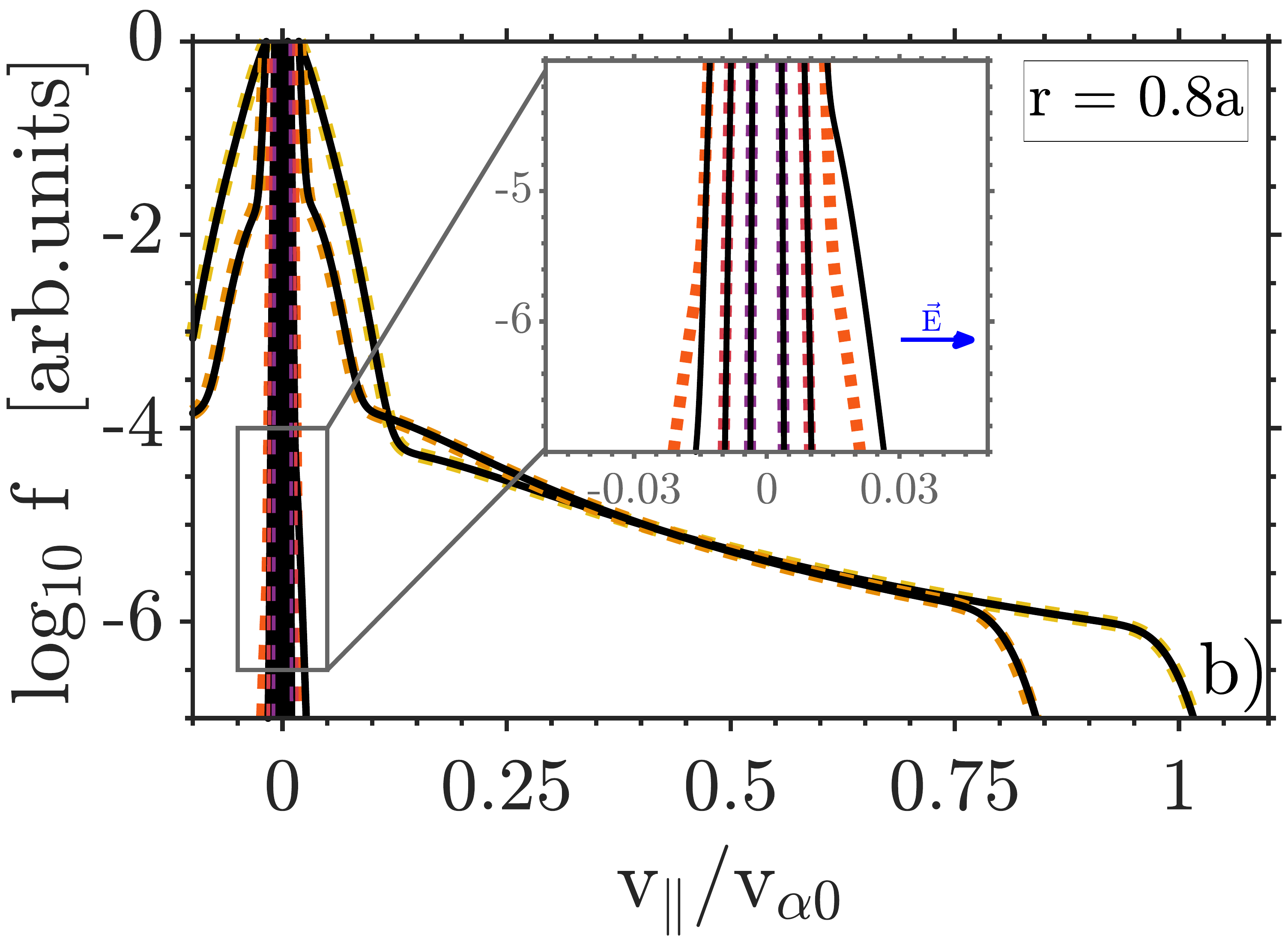}
\includegraphics[width = 0.48\linewidth]{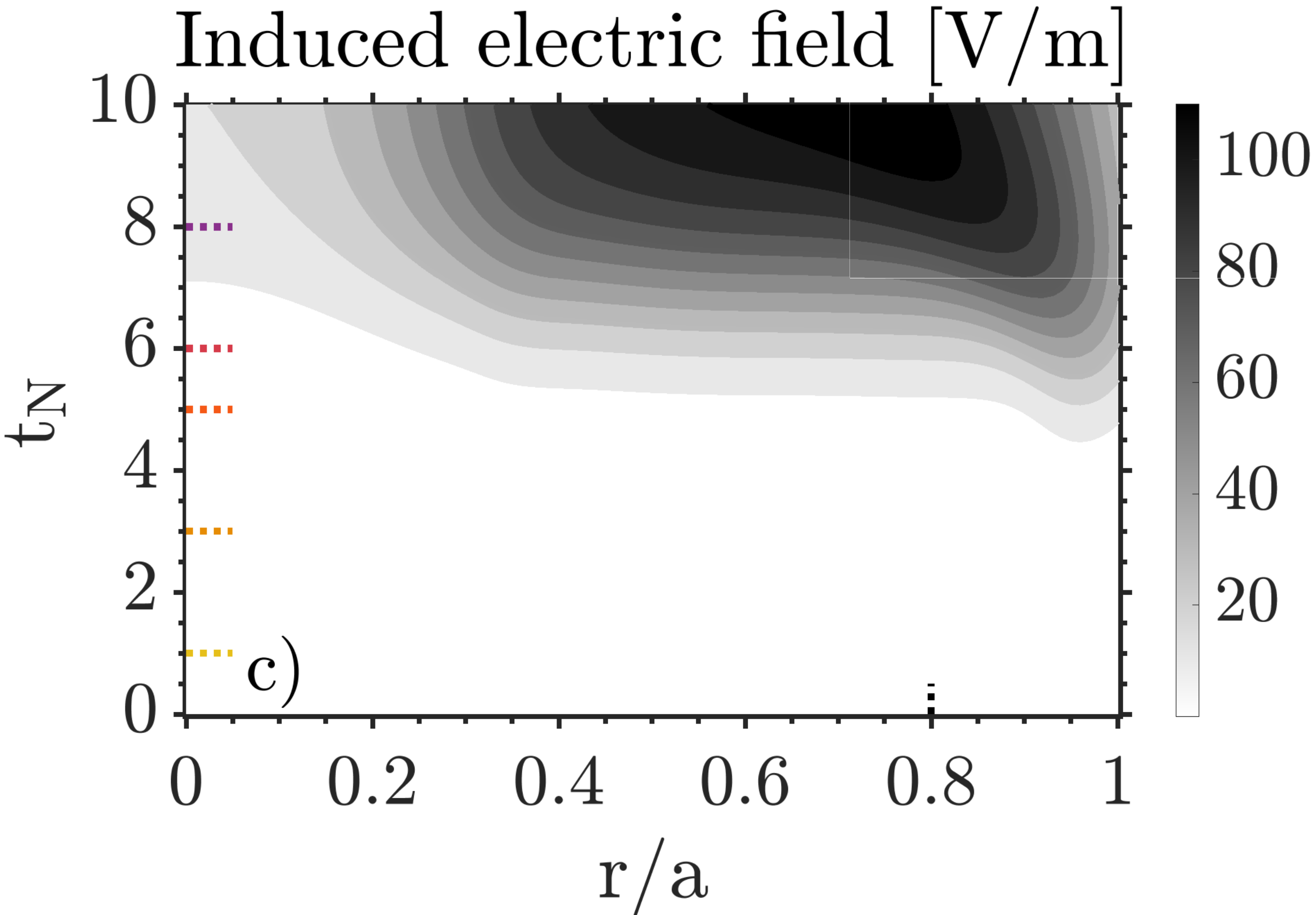}
\caption{a) The alpha velocity distribution $f(v,t)$ in the ITER core plasma during a thermal quench, evolving without (dashed, color) and with (solid, black) electric field input. b) The same simulation as in a) but at $r/a =0.8$. The colors indicate the time steps and are the same as in a). Figure c) shows the induced electric field that is used as input for a) and b). The dashed lines mark the time-points (color) and radial position (black) used in a) and b).}
\label{fig:codion_go}
\end{figure}

We turn to an unmitigated case ($n_{e1} = n_{e0}$) with $t_\mathrm{TQ}=1\unit{ms}$. Using global temperature and density profiles of the 15~MA D-T plasma scenario (figure~\ref{fig:pre_dis}a, details in section~\ref{sec:equil}) we can calculate the induced electric field using the fluid-code GO~\cite{feher11simulation,papp13effect,vallhagen20REs}. It solves the induction equation in 1D with the effects of radial electric field diffusion included. The evolution of the electric field is displayed in figure~\ref{fig:codion_go}c as a function of the normalized radius $r/a$ with $a=2.06\unit{m}$ minor radius. The electric field becomes significant at roughly $5 t_N$. With CODION we calculate the evolution of $f(v,t)$ at two spatial points $r=0$ and $r=0.8a$ and show the results in figure~\ref{fig:codion_go}a and \ref{fig:codion_go}b both with and without $E$-field input. The electric field accelerates ions in positive parallel velocity $v_\|$ direction and is shown to have negligible effect on the distribution $f$. Even when the electric field reaches significant amplitudes, the collisional slowing-down due to the still cooling background overcomes the acceleration. The kinetic alpha particle pressure $p_\alpha$ (eq.~(\ref{eq:p})) changes on a sub-percentage level when comparing the simulations with and without $E$-field. Even in a prolonged CODION simulation ($20 t_N$) the electric field is not able to drag out a tail of energetic ions. This is in agreement with previous studies~\cite{embreus15numerical}, finding ion runaway to be unlikely on tokamak disruption time-scales. 

The alpha particle calculations reside within the TQ, whose temperature evolution for this work we assume to be dominated by MHD losses~\cite{smith08hot}, i.e. less sensitive to material composition. As such, the delay in the electric field induction is independent of $n_{e1}$ and $Z_1$ and remains the same on a $t_N$ (temperature decay) time scale. On the other hand, the alpha thermalization of the unmitigated case is the slowest (figure~\ref{fig:dis_val}b). Thus we conclude, that for the thermal quenches considered in this work ($t_\mathrm{TQ} \geq 1\unit{ms}$) the velocity space isotropy demonstrated explicitly for the unmitigated case here can be assumed for mitigated cases as well. 

\newpage
\section{Alpha particle interaction with Alfv\'enic modes in mitigated ITER thermal quenches} \label{sec:alfven}
We aim to determine the interaction between energetic alpha particles and waves in the plasma. Theoretical predictions~\cite{pinches15ions} show that the instability drive coming from alpha particles in an ITER plasma are on par with background damping in the hot D-T phase. The damping however originates from the bulk plasma and its efficiency drops dramatically during the TQ~\cite{lier21alpha}. Together with the knowledge that alpha particles resist the thermalization for a significant amount of time (see section~\ref{sec:coll_alpha}), there is an opportunity for modes to be driven unstable during the thermal quench of a disrupting plasma. Investigating such a mechanism requires the calculation of wave-particle interactions, which in turn requires obtaining the alpha particle distributions in both real space and velocity space, as well as a calculation of plasma equilibrium, mode spectrum and mode damping, which are the subject of this section.

\subsection{The spectrum of weakly damped Toroidal Alfv\'en Eigenmodes} \label{sec:equil}
\begin{table} [b]
\centering
\begin{tabular}{lcc}
Parameter name&Notation&Value\\ \hline
Major radius & $R_0$ & 6.195 m\\
Minor radius & $a$ & 2.06 m\\
Effective charge & $Z_\mathrm{eff}$ & $\approx 1.0$\\
Normalised flux & $\psi$ & $\Psi(r)/\Psi(a)$\\
Normalised radius & $r/a$ & $r/a\simeq\sqrt{\psi} \equiv s$\\
Plasma current & $I_{p0}$ & 15 MA\\
Magnetic field on axis & $B(s=0)$ & 5.26 T\\
Electron density on axis & $n_{e0}(s=0)$ & $10^{20}$ m$^{-3}$\\
D-T density on axis & $n_\text{DT0}(s=0)$ & $0.5\cdot 10^{20}$ m$^{-3}$\\
Electron temperature on axis & $T_{e0}(s=0)$ & 24.7 keV\\
Ion temperature on axis & $T_{i0}(s=0)$ & 21.2 keV\\
\end{tabular}
\caption{\label{tab:parameters}The main plasma parameters of the 15~MA ITER scenario. Note that for models that are written in flux coordinates, the transformation $r/a \approx s$ is done.}
\end{table}

\begin{figure}
\centering
\includegraphics[width = 0.48\linewidth]{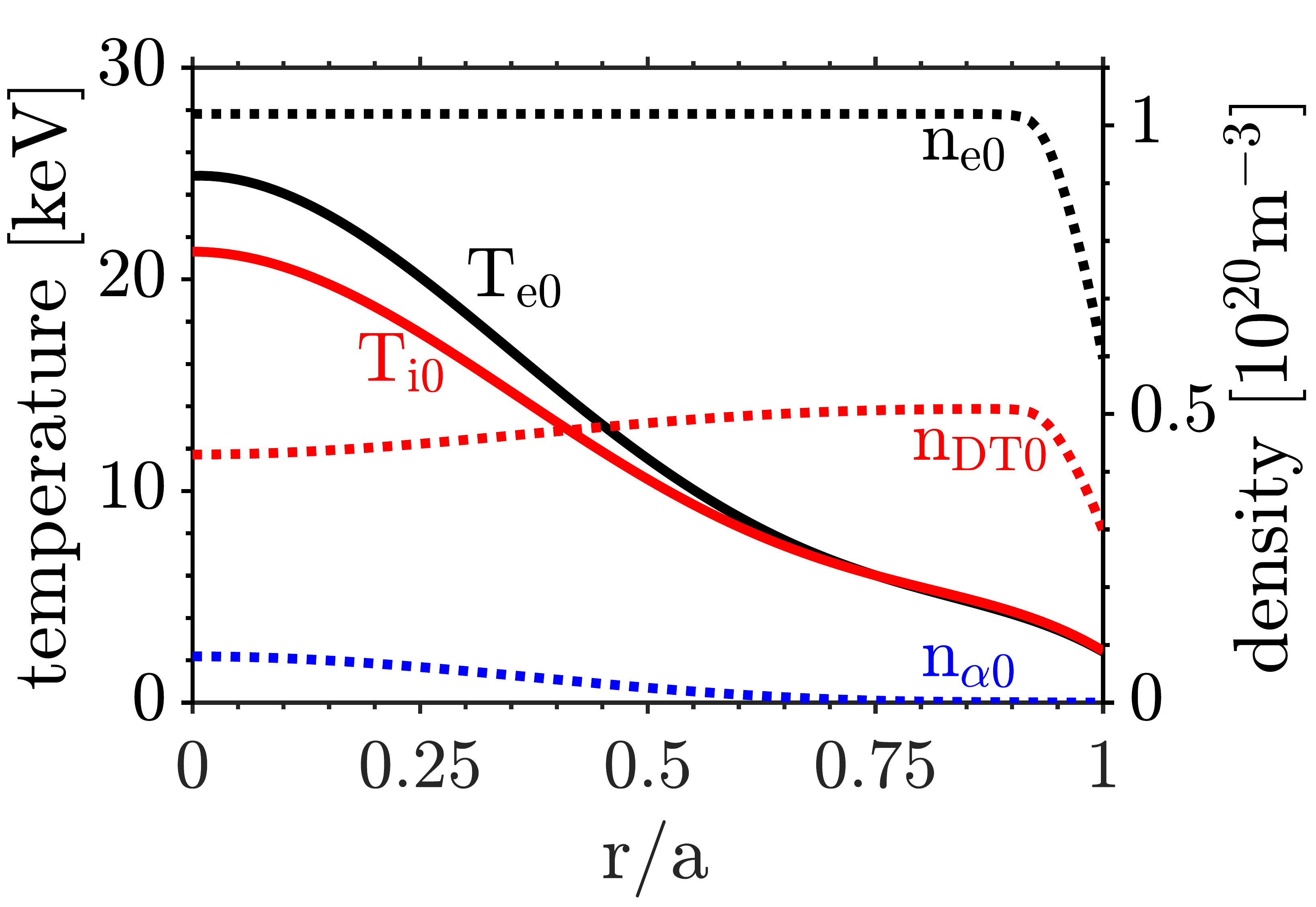}
\caption{Pre-disruption profiles of densities and temperatures of the D-T ITER plasma phase according to ``scenario \#2'' for a case without impurities.}
\label{fig:pre_dis}
\end{figure}

The modelling begins with the 15~MA inductive D-T plasma ``scenario \#2'' described by Polevoi {et al.}~\cite{polevoi02iter,iter_scen2}, which determines the pre-disruption plasma conditions. Core-parameters ($r/a = 0$) correspond to the values validated against in section~\ref{sec:model} and the temperature and density profiles are shown in figure~\ref{fig:pre_dis}. The ion composition consists of equal deuterium and tritium densities $n_{DT0}$ and the alpha particle minority $n_{\alpha 0}$, 
related to $n_{e0}$ through quasi-neutrality $n_{e0} = 2 n_{DT0} + Z_\alpha n_{\alpha 0}$. The main plasma parameters are given in table~\ref{tab:parameters}. A radial grid of 101 points is set up and populated each with a steady-state slowing-down distribution $f_0 (r,v)$ (eq.~(\ref{eq:fsd})) according to the temperature and densities of the operation scenario. Other impurities are not included here. 

The thermal energy loss is prescribed with an exponential decay for profiles of temperature:
\begin{equation}
T_1 (r,t) = T_f + \left[ T_{e0} (r) - T_f \right]\mathrm{e}^{-t_N}, \label{eq:T}
\end{equation}
and the plasma composition changes as a step-function increase in density (eq.~(\ref{eq:ne1})). The calculation of the mode spectrum is conducted for the thermal quench, whose temperature evolution we assume to be MHD-dominated. In reality, material injection and temperature evolution are not independent of each other, but determining the exact relationship is outside the scope of this work.

\begin{figure}
\includegraphics[width = 0.48\linewidth]{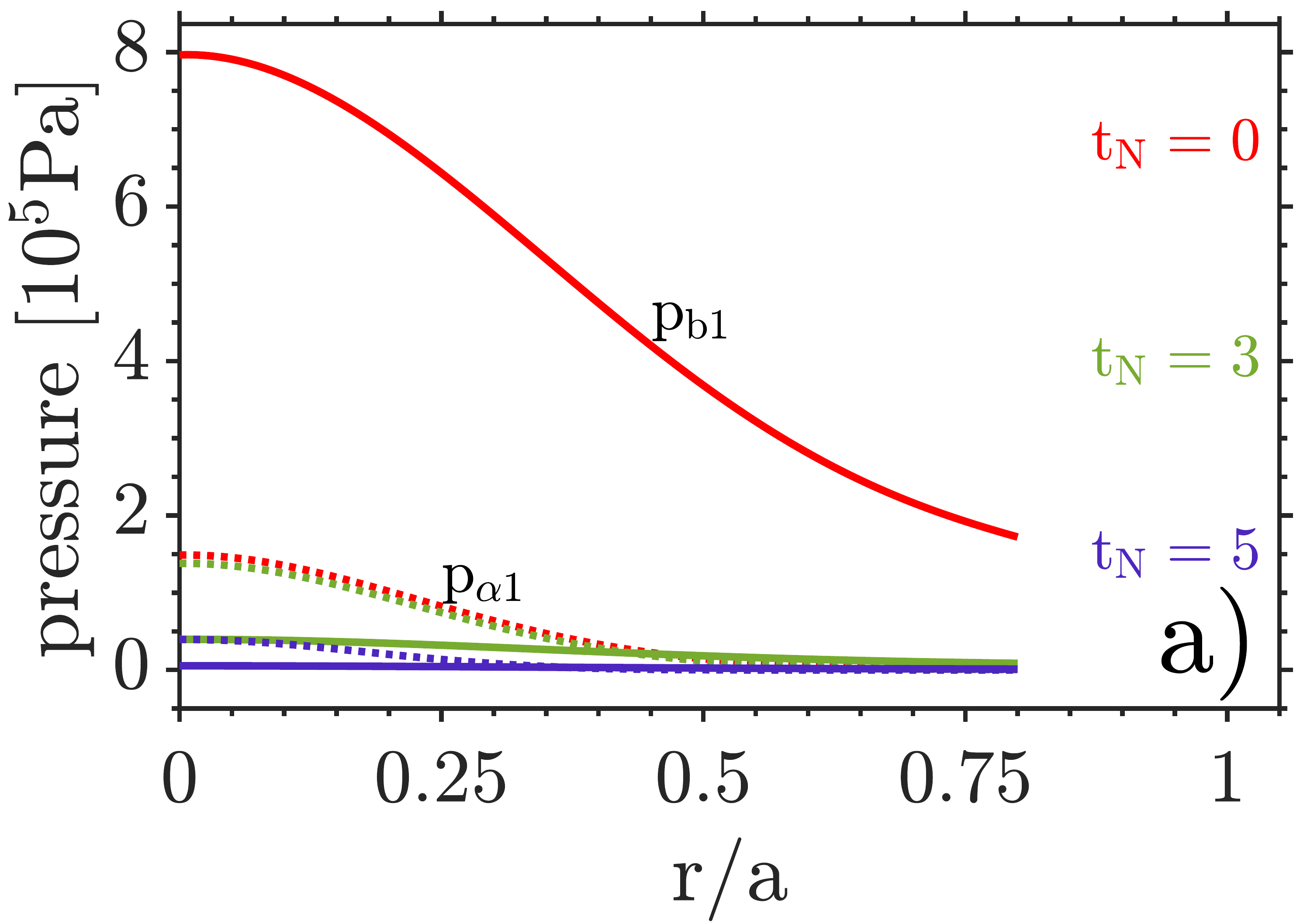}
\includegraphics[width = 0.48\linewidth]{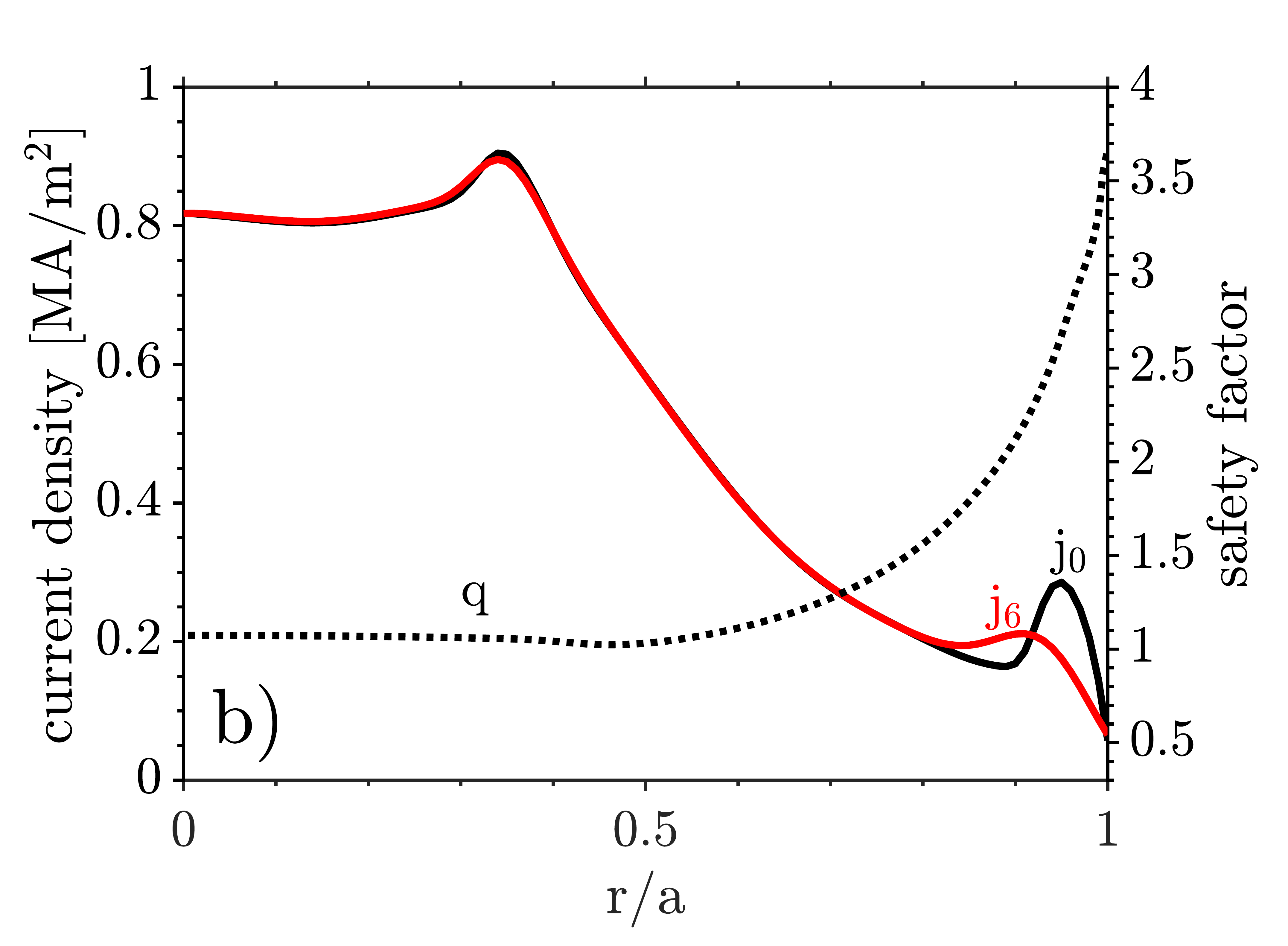}
\caption{a) Time evolution of the background pressure $p_{b1}$ (solid) and the alpha pressure $p_{\alpha 1}$ as a function of normalized time $t_N = t/t_\mathrm{TQ}$ and indicated by color. b) Calculated safety factor profile $q$ and current density profiles for $t_N= 0$ ($j_0$) and $t_N =6$ ($j_6$) as obtained by the fluid code GO~\cite{feher11simulation,papp13effect,vallhagen20REs} and used for the plasma equilibrium reconstruction.}
\label{fig:p_j}
\end{figure}

A plasma equilibrium can be described by magnetic flux surfaces of constant pressure and calculated via the Grad-Shafranov equation~\cite{wesson04tokamaks}, requiring the input of profiles of current density and pressure. A plasma equilibrium valid for the thermal quench was reconstructed (using VMEC~\cite{VMEC}) for the preceding study on unmitigated disruptions~\cite{lier21alpha}. The pressure-profile used for the reconstruction consists of a thermal background pressure $p_1 = 2 n_{e0} T_1 (r,t_N = 3)$ and an alpha particle pressure obtained from CODION. The pressure in CODION is calculated from the moment of the numerically calculated distribution function, see eq.~(\ref{eq:p}). Though alpha particles resist the sudden drop in pressure for a few milliseconds, the level of total pressure remains low compared to the pre-disruption condition (figure~\ref{fig:p_j}a). On the other hand, the current density $j$ is barely changing during the TQ as depicted in figure~\ref{fig:p_j}b for $t_N = 0$ ($j_0$) and $t_N=6$ ($j_6$). With the thermal pressure exponentially decaying, $j_0 \simeq j_6$ becomes the dominating factor in the equilibrium reconstruction. The $q$-profile of the equilibrium is shown in figure~\ref{fig:p_j}b. It has an on-axis value of 1.071, a local minimum at $r/a \approx 0.5$ and $q > 1$ holds throughout the entire plasma. In order to account for deviations in the disruption scenario, a shape-preserving scan over the elevation of the safety factor profile was conducted~\cite{lier21alpha}. The sensitivity measure hereby is the spatial location and number density of Alfv\'enic modes, that are going to be used for further calculation. The scan showed a wide availability of frequency gaps and toroidal Alfv\'en Eigenmodes (TAEs) irrespective of the absolute value chosen within the scan. For this study we assume, that this previously reconstructed equilibrium is not significantly altered by material injection on the timescale considered. The usage of the equilibrium in this study does not go beyond $t_N =6$.

\begin{figure}
\includegraphics[width = 0.48\linewidth]{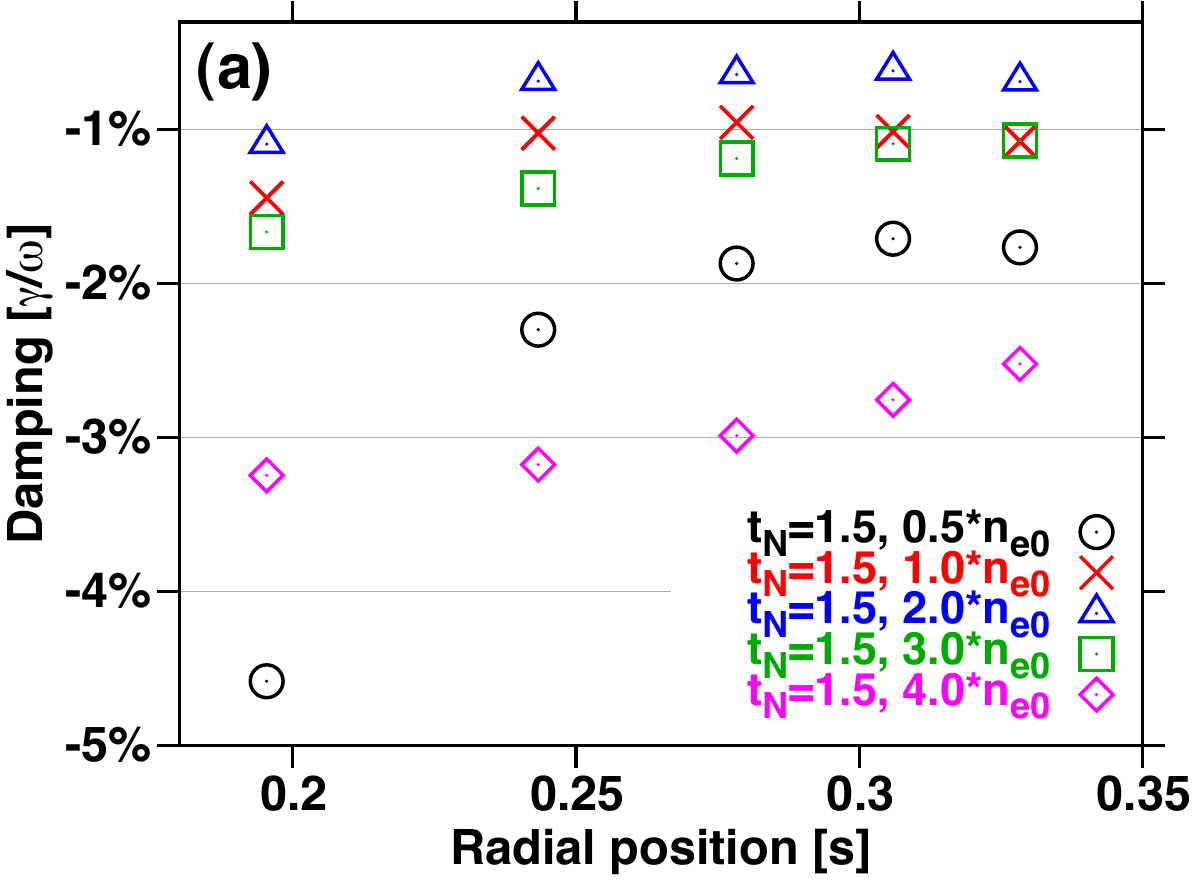}
\includegraphics[width = 0.48\linewidth]{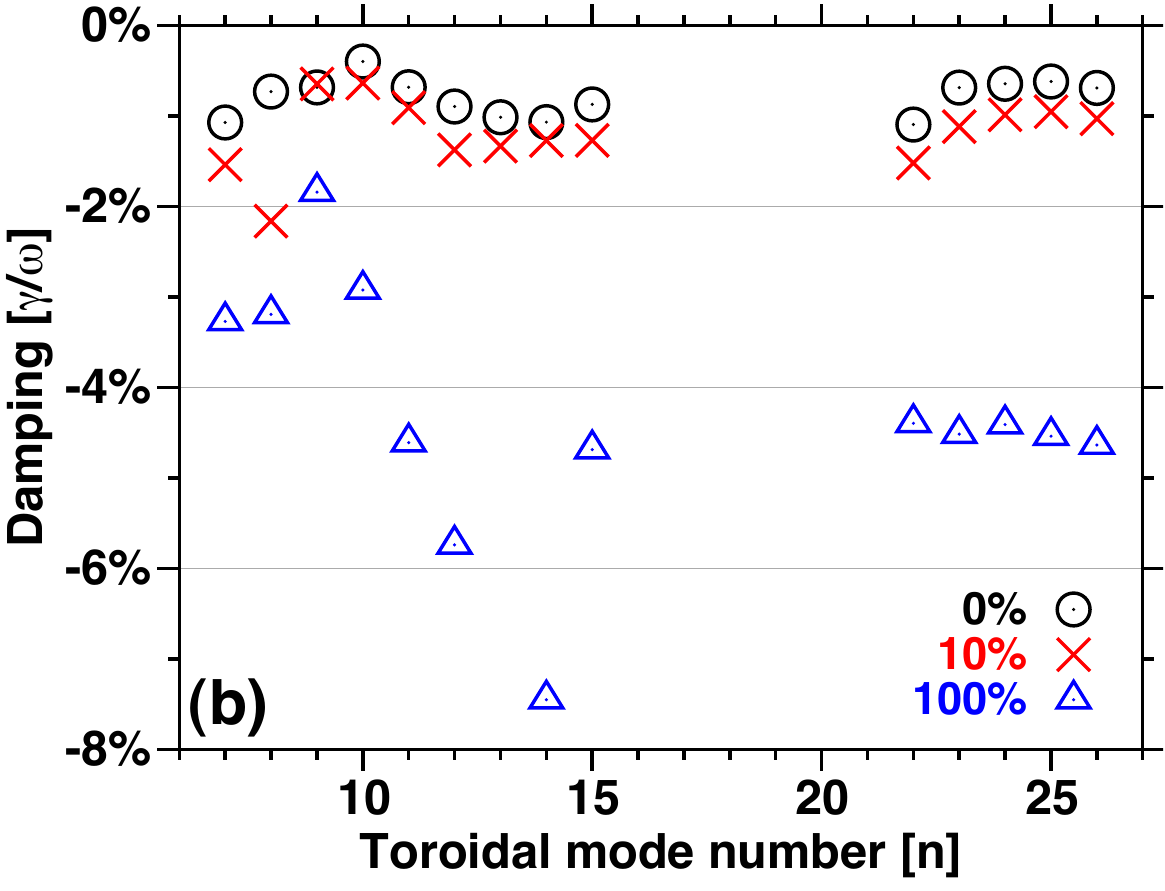}
\caption{a) Damping rates of the high-$n$ branch at $t_N= 1.5$ and as a function of the initial electron density $n_{e0}$. The low-$n$ branch shows the same trend but is not depicted. b) The damping of both branches is shown as a function of injected neon density in \% for a post-disruption density of $n_{e1} = 2 n_{e0}$ and for $T_1(t_N=1.5)$.}
\label{fig:damping2}
\end{figure}

With the use of the linear gyrokinetic code LIGKA~\cite{LIGKA,LIGKA2} we search for toriodicity-induced frequency gaps in the ideal MHD spectrum of the equilibrium. This reveals Toroidal Alfv\'en Eigenmodes (TAEs) that lie within Alfv\'en continuum gaps and are therefore weakly damped. The TAEs are represented in Fourier-space, with toroidal mode numbers $n$ and poloidal mode numbers $m$ and are located around the radial position $r_\mathrm{TAE}$, where $q$ fulfils
\begin{equation*}
q(r_\mathrm{TAE}) = \frac{ m+1/2}{n}. 
\end{equation*}
In addition, LIGKA calculates individual mode damping rates, including nonlocal continuum damping~\cite{tataronis73mixing}, ion/electron Landau damping~\cite{landau46damping1,candy96eldamping,fu89excitation} and radiative damping~\cite{mett92damping}. The collisional damping on trapped electrons, and resistive fluid damping were calculated in the preceding study~\cite{lier21alpha} and deemed insignificant up to a global time of $t_N = 8$.
LIGKA is written in the ``PEST'' coordinates~\cite{grim83pest}, thus the transformation $r/a \rightarrow s$ takes place. Before calculating the actual wave-particle interaction, we investigate the behaviour of damping as a function of the evolving background temperature and plasma densities. For the damping calculations the alpha particle presence is neglected, choosing $Z_\text{eff}\approx 1$.  Small amount of impurities change the damping only slightly~\cite{pinches15ions}. At higher impurity contents (following MMI) the Alfv\'en velocity changes, leading to a dramatic change in the resonance and leading to increased damping. The effect of impurities on damping will be discussed in detail later in this section (figure~\ref{fig:damping2}b). Changes to the safety factor (within the bounds of the scan conducted~\cite{lier21alpha}) have no major impact on the damping.

\begin{figure}
\includegraphics[width = 0.49\linewidth]{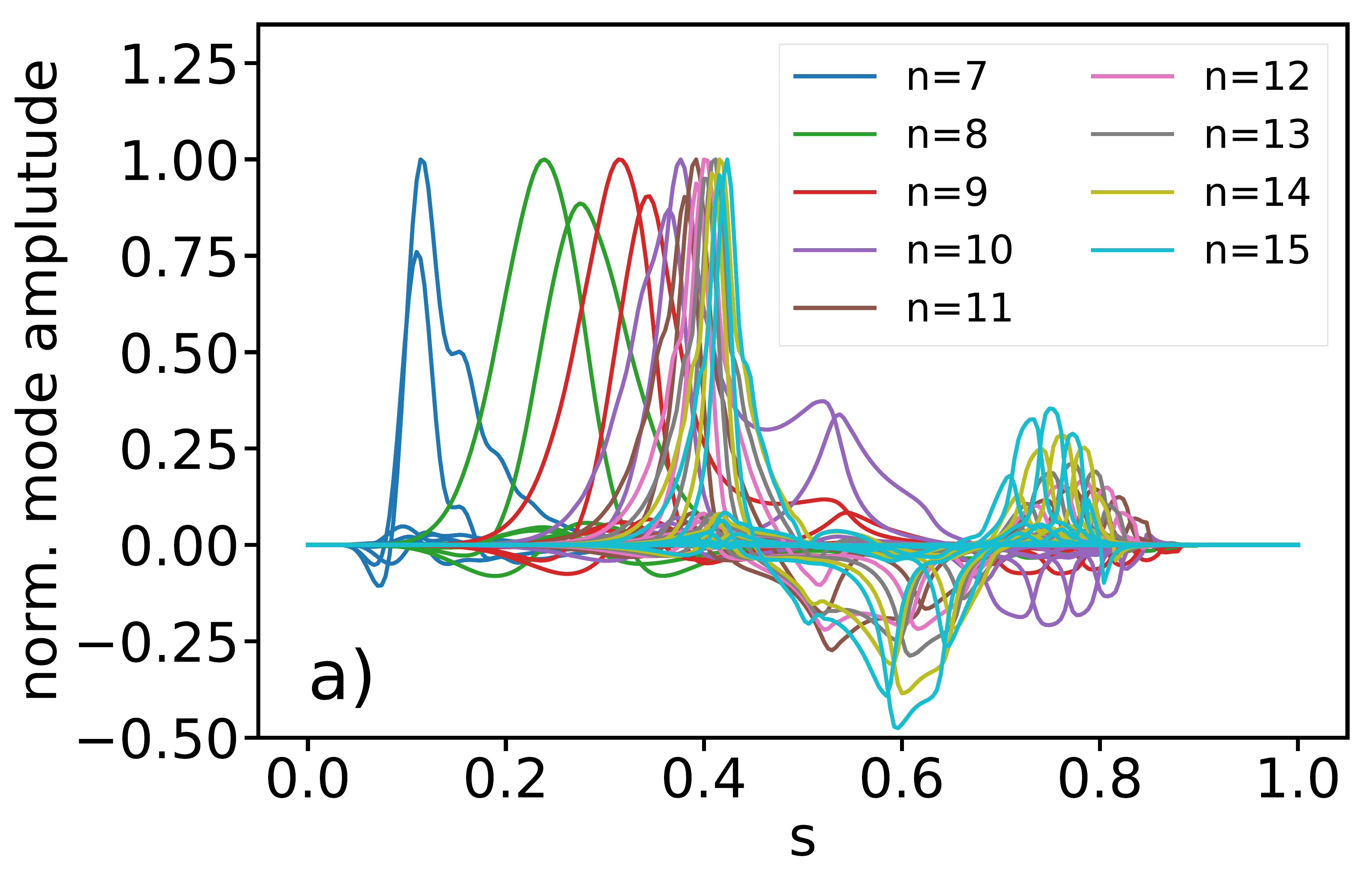} 
\includegraphics[width = 0.49\linewidth]{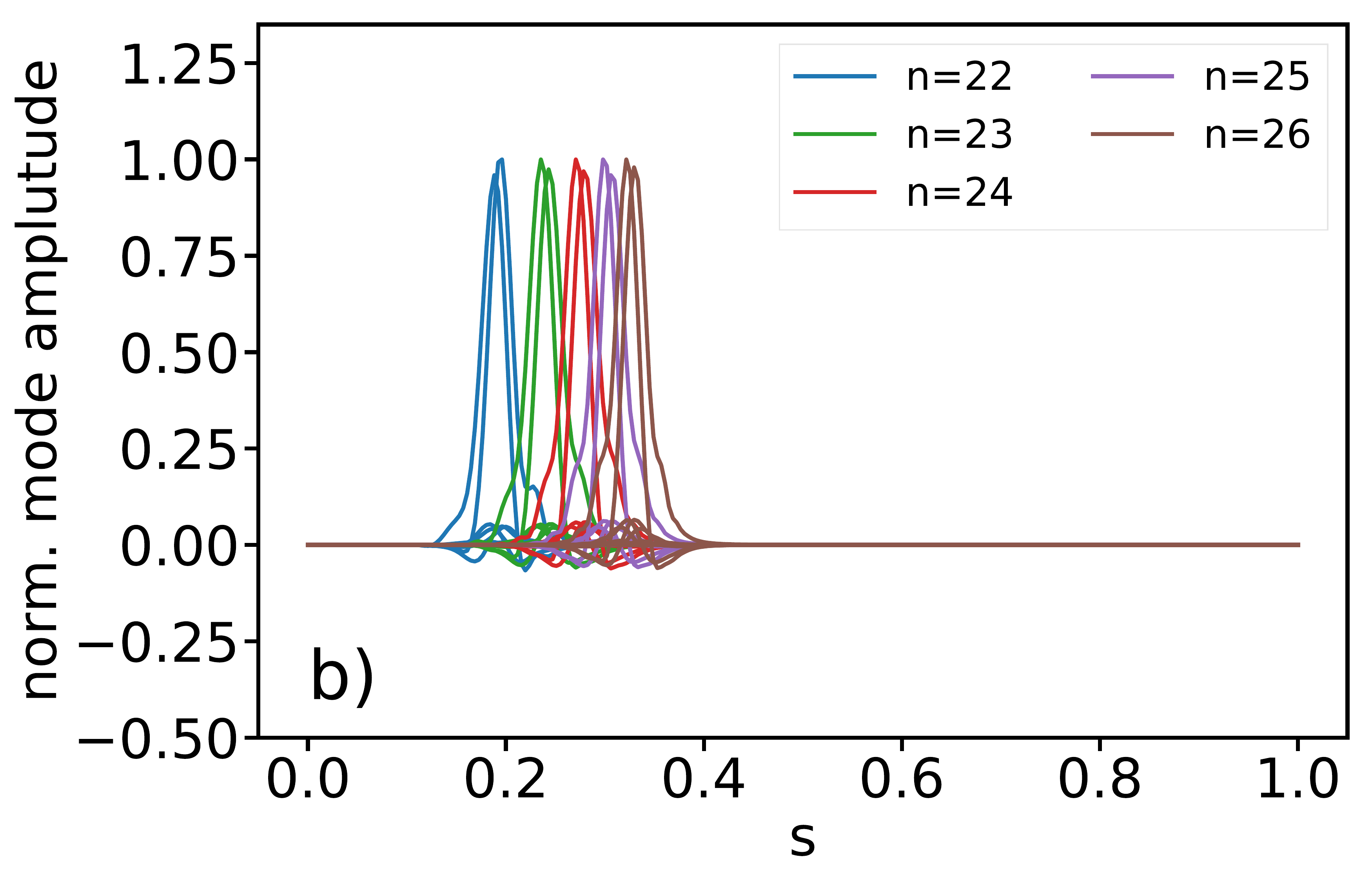} 
\caption{Figures of the normalized eigenfunctions of the TAEs found in the post-disruption equilibrium for a) the low-$n$ branch and b) the high-$n$ branch. All the individual poloidal harmonics of the electrostatic potential are included. The eigenmodes shown are of even parity, as seen by the equal sign of the dominating poloidal harmonics.}
\label{fig:mode_structures}
\end{figure}

For now, the calculations are restricted to the inner half of the plasma ($s < 0.5$) and to TAEs with even parity (with respect to the poloidal angle), yielding a set of modes that will be denoted with $M_1$. The restriction is motivated by the alpha particle spatial location (see figure~\ref{fig:p_j}a), whose spatial pressure gradient will ultimately be the driving force for the modes.  Global mode structures are shown in figure~\ref{fig:mode_structures}. Generally, we find TAEs with a low-$n$ branch ($n=7-15$) and a high-$n$ branch ($n=22-26$). Up to 13 poloidal harmonics are used for the representation in Fourier space and their frequencies range from $74-83\unit{kHz}$. Because of the flatness of the $q$-profile in the inner half of the plasma, a high density of neighbouring TAEs with spatial overlap is found. For wave-particle interactions, this promises a resonance overlap in the phase-space and can cause particle transport~\cite{schneller16alfven}.

\begin{figure}
\includegraphics[width = 0.48\linewidth]{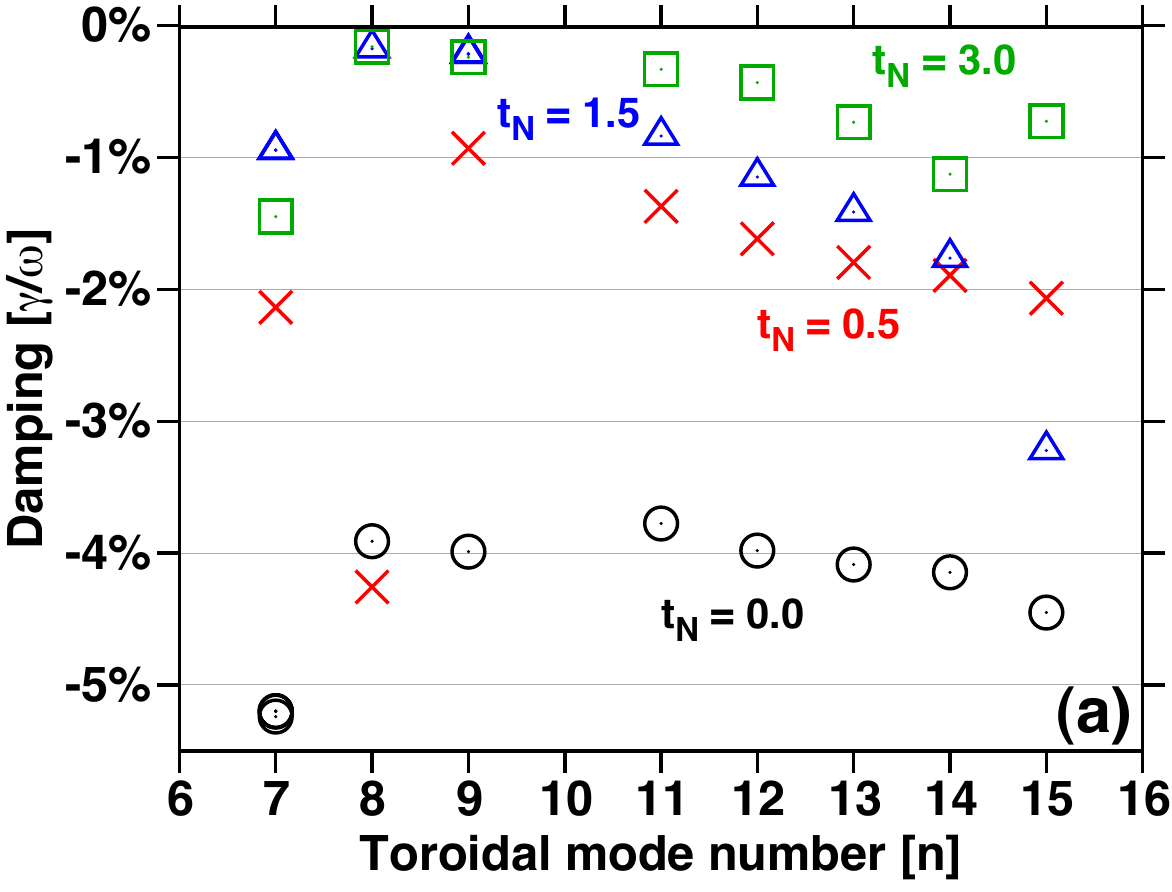}
\includegraphics[width = 0.48\linewidth]{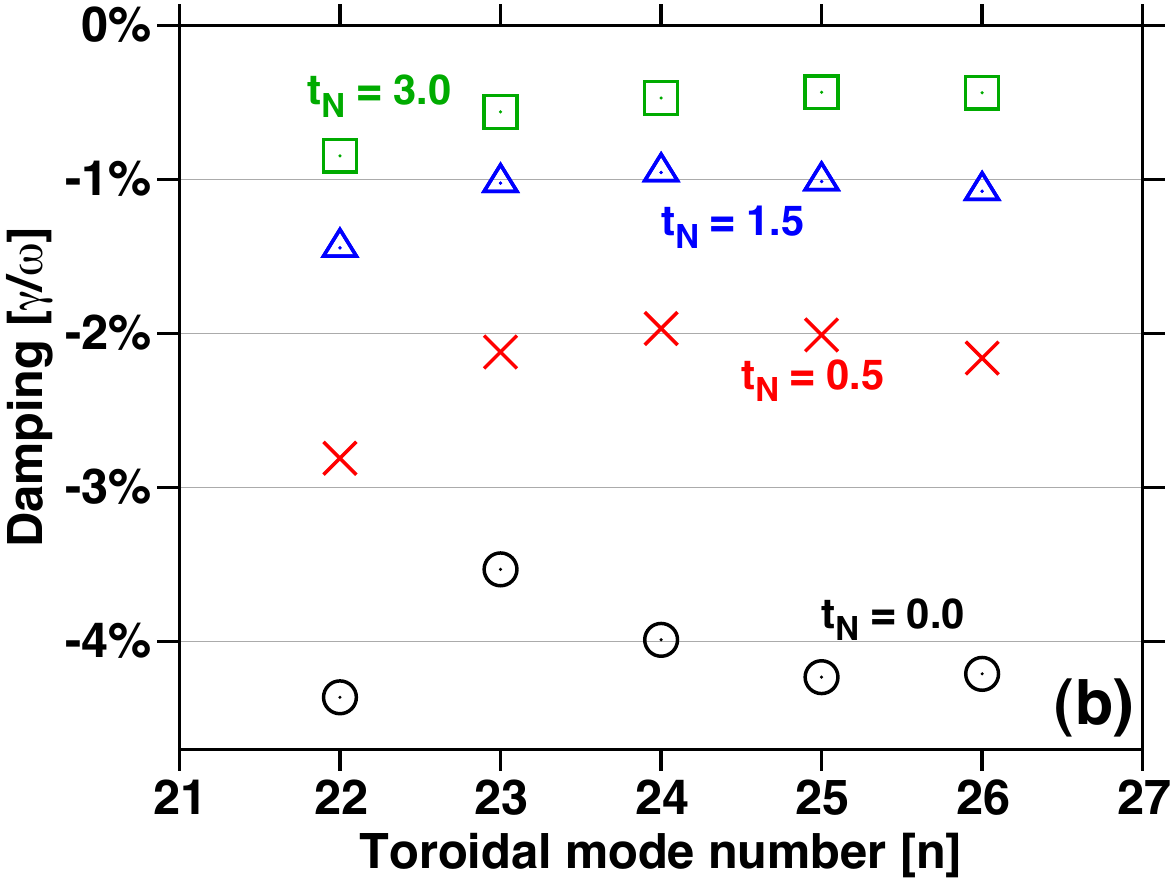}
\caption{Damping over time $t_N$ of TAEs at the a) low-$n$ and b) high-$n$ branch.}
\label{fig:damping}
\end{figure}

LIGKA runs are conducted for $n_{e1}= n_{e0}$ and time-evolving temperature profiles, which show that the total damping initially decreases as a function of $t_N$ (figure~\ref{fig:damping}). The dominant ion Landau damping mechanism of a hot plasma is based around Maxwellian tail ions resonating with the wave at $(v_A, v_A/3)$~\cite{heidbrink08basic}, where $v_A \equiv B/\sqrt{\mu_0 \rho} \gg v_{ti,\|}$ is the Alfv\'en velocity, with the vacuum permeability $\mu_0$, the plasma mass density $\rho$ and the parallel thermal ion velocity $v_{ti,\|}$. It is expected to be the most significant damping mechanism for D-T plasmas, but due to the resonance with the Maxwellian tail it is exponentially sensitive to the ion temperature~\cite{pinches96phd,pinches15ions}. As for the parallel thermal electron velocity $v_{te,\|} \gg v_A$ holds, only a small portion of the electron distribution can partake in damping. It is known~\cite{candy96eldamping}, that the electron Landau damping evolves proportional to the electron pressure $\beta_e \propto T_1$ and therefore decays accordingly. The continuum damping for TAEs is essentially zero and therefore unaffected by the temperature evolution. Radiative damping however is related to finite Larmor radius effects. As the Larmor radius shrinks during the thermal quench the radiative damping loses effectiveness as well. When plasma temperatures reach orders of electronvolts, damping mechanisms of a cold plasma need to be addressed.

The above LIGKA simulations are repeated for various injection amounts of deuterium and neon mimicking the effects of MMI systems. In figure~\ref{fig:damping2} we show the effects of density changes at $T_1(t_N=1.5)$ onto the damping of the high-$n$ branch, though effects on the low-$n$ branch are similar. The material injected is modelled as a Maxwellian distribution at a temperature of 10~eV and deposited equally throughout the plasma. While resonant effects with 10~eV Maxwellian and harmonics of $v_A$ are unlikely, the injection changes the charged particle mass and causes the Alfv\'en velocity to shift relative to the thermal velocities of the main populations. The combined evolution of ion and electron Landau damping leads to the observed changes in the TAE damping rates~\cite{lier21alpha}. We conduct additional LIGKA simulations with changes to the injected material composition, which now consists of 10\% and 100\% (singly ionized) neon, rest deuterium. The heavy neon population has a significant effect onto the damping strength even for modest injection amounts $n_{e1} = 2n_{e0}$, as shown in figure~\ref{fig:damping2}b.

\newpage
\subsection{Wave-particle interaction} \label{sec:wp_int}
Competing against the background damping is the mode drive from the energetic alpha particle distribution~\cite{heidbrink08basic}. Analytical estimates of the EP drive exist~\cite{fu89excitation} but we will compute the wave-particle interaction more precisely with HAGIS~\cite{pinches04role} - a perturbative, non-linear code. The tool calculates mode evolution in the presence of energetic particles as well as the redistribution of the particles that is caused by the non-linear interaction. Since the radial pressure gradient is determining mode drive, a transport model for the alpha particles will be added in the form of a diffusion. We begin however with a simulation on an unmitigated disruption case with $t_{\rm TQ}=1\unit{ms}$ and without transport. This is further extended to a parameter space evaluation in order to discuss the effects of alpha particle transport, thermal quench time and material injection.

\subsubsection{Alpha-driven TAE mode evolution during an unmitigated ITER plasma disruption}

The HAGIS simulation requires an input of the post-disruption equilibrium, the modes obtained by LIGKA (section~\ref{sec:equil}) and the alpha particle distributions. Utilizing the model presented in section~\ref{sec:coll_alpha}, we populate the plasma with distribution functions $f(v,r,t_N=1.5)$ according to plasma parameter profiles and provide them analytically to HAGIS. The LIGKA eigenmodes $M_1$ are imported and set to an initial mode amplitude (relative to the on-axis guide field) of $\delta B/B = 10^{-10}$. We choose $t_N=1.5$ as the beginning time-point for the HAGIS simulations because of the low damping calculated (see figure~\ref{fig:damping}). The Alfv\'en velocity for the unmitigated ITER plasma is $v_A/v_{\alpha 0} \approx 0.58$ with the most fundamental resonances occuring at $(v_A, v_A/3)$~\cite{heidbrink08basic}. As shown in figure~\ref{fig:dis_val}a,  figure~\ref{fig:codion_go}a and \ref{fig:codion_go}b, this region of the velocity space is well populated by energetic alphas at $t_N=1.5$.

The alphas are represented by $10^5$ markers and the integration time-step in the mode evolution is $5 \cdot10^{-7}\unit{s}$. The simulation duration is limited to the point of complete slow-down of the alphas (at $s=0$), but we will see that the growth is strong enough for the modes to saturate well before the end of the simulation. In this time frame the particles are redistributed by HAGIS through their interaction with the TAEs. Mode damping as calculated by LIGKA is taken as constant, as are the phases and structures of the modes.  

\begin{figure}
\includegraphics[width = 0.48\linewidth]{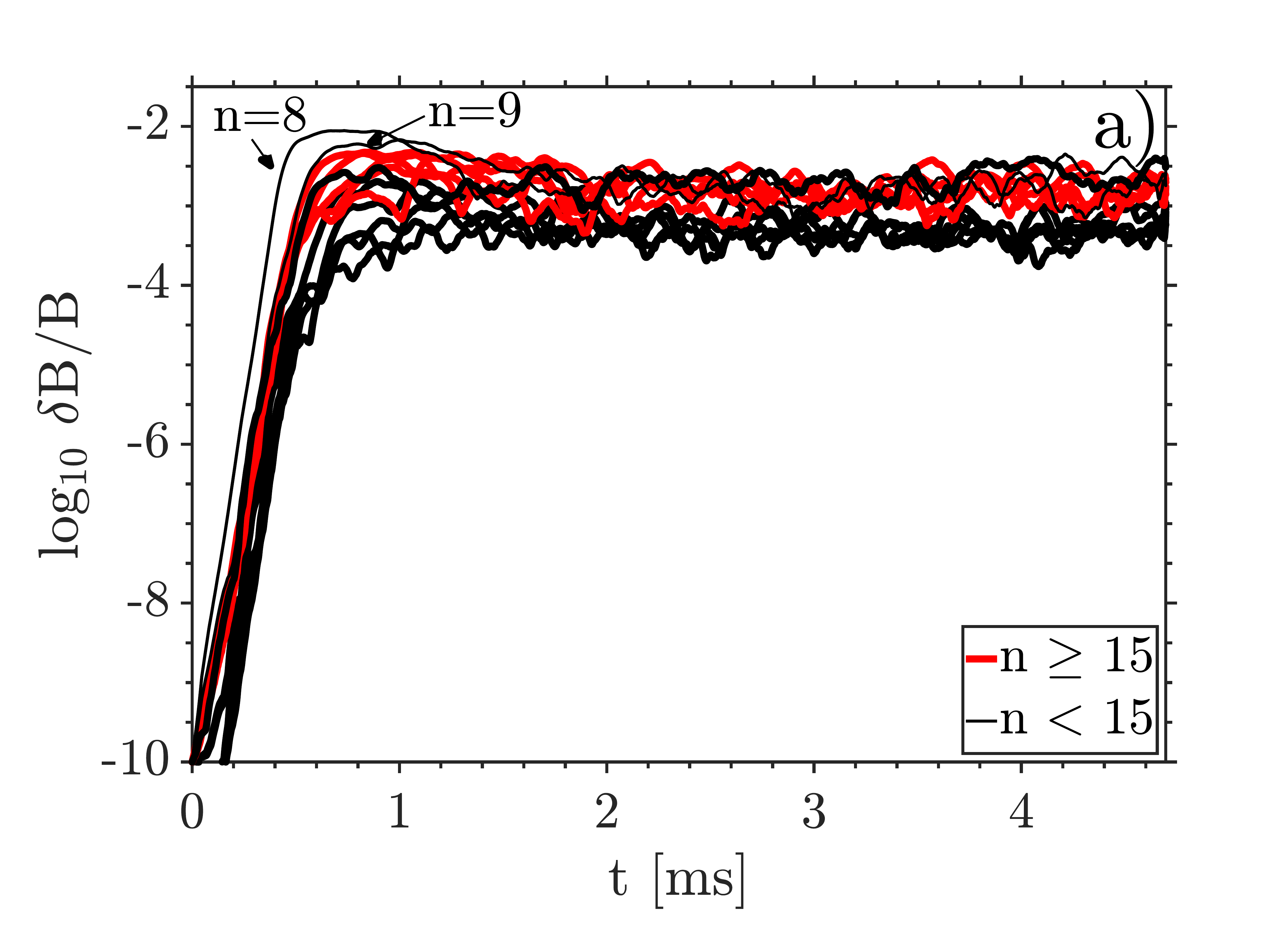}
\includegraphics[width = 0.48\linewidth]{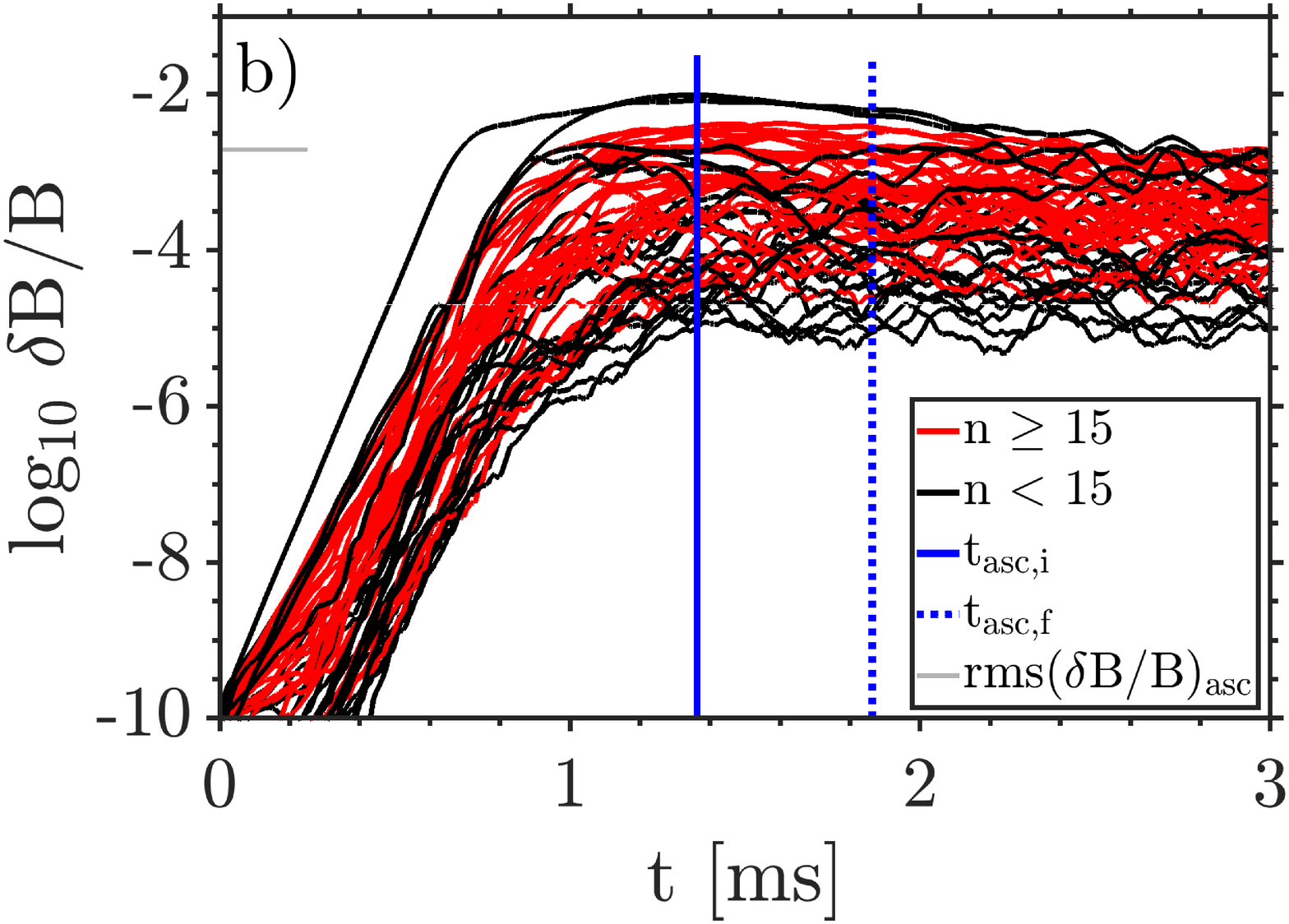}
\caption{Mode evolution of TAEs in a post-disruption ITER plasma for the unmitigated disruption case ($t_{\rm TQ} = 1\unit{ms}$). Figure a) shows the simulation with only the modes present in the inner half of the plasma ($M_1$), while the b) shows the same simulation with the entire plasma taken into account ($M_2$). The end of the simulation corresponds to the time point of the alphas having slowed down $v_\alpha (s=0) = 0$. Legend entries with subscript ``asc'' in b) refer to its further usage in section~\ref{sec:dream}. The displayed data has been slightly smoothed for the purpose of illustration.}
\label{fig:evo}
\end{figure}

In figure~\ref{fig:evo}a we show the mode evolution conducted with the inner set of modes $M_1$. We see a strong linear growth phase with $\gamma/\omega \approx 1.8\%$ and a saturation at approximately 1~ms. Particularly standing out are the $n=8$ and $n=9$ modes, which have the lowest (essentially zero) damping rates (see figure~\ref{fig:damping}a). They briefly reach amplitudes of $\delta B/B \leq 1\%$, meaning the result has to be taken with caution. It is known~\cite{schneller13alfv} that due to a lack of zonal-flow physics and mode-mode coupling effects, the HAGIS model can overestimate mode amplitudes. As such, the mode amplitudes are treated as an upper limit. Lower values are obtained with a larger set of modes: For computational reasons (and because the damping/alpha-drive in the radial direction outwards increases) we restricted the Eigenmode searcher in the LIGKA tool to the inner half of the plasma and to even parity TAEs. We now lift these restrictions and repeat the HAGIS simulation with a new set of modes, $M_2$, that includes all the relevant TAEs in the plasma. The $M_2$ set of modes has toroidal mode numbers $n = 6-26$, with some gaps populated with more than one TAE (even and odd), totalling in 62 toroidally distinct modes. The mode evolution is shown in figure~\ref{fig:evo}b, showing the (even) $n=8$ to still be the strongest driven mode, however with a slightly reduced growth rate. Part of the low-$n$ TAE branch (black) now saturates at $\delta B/B \approx 10^{-5}$ and consists of odd parity TAEs that receive generally less EP drive due to their higher frequency $f \sim 100\unit{kHz}$. With up to 17 poloidal modes and a higher resolution requirement, the $M_2$ set is computationally demanding on the HAGIS code. It will be used for a self-consistent calculation of a disruption plasma in section~\ref{sec:dream}, however, the broad parameter scan of the next section will be calculated using $M_1$.

\subsubsection{Alpha-driven TAE mode evolution in mitigated ITER plasma disruptions and the effects of alpha particle diffusion}

Previous wave-particle interaction calculations with the HAGIS code are extended to a parameter space $\mathbf{A} = [D_\alpha, t_\mathrm{TQ}, n_{e1}, n_\text{Ne1}/n_{D1}]$, where we add a yet to be defined diffusion parameter $D_\alpha$ in addition to the resonant EP transport in the HAGIS model.

Disruption-induced transport of particles (esp. for ITER) is a subject hard to assess without costly numerical simulations. A plasma disruption is regularly accompanied by the breakup of the nested magnetic flux surfaces~\cite{boozer12disruptions}. The particle transport - especially for high-velocity particles like alphas -  is influenced mainly by the healing rate of those surfaces. In order to avoid costly numerical calculations, this problem will instead be addressed with a diffusion model for the alphas, which allows for parameter scans to be conducted and sensitivities to be explored.

We treat the transport as a one-dimensional diffusive process and separate the real space problem from our velocity space solution. We state the one dimensional diffusion equation for the particle density $n_\alpha(r,t)$:
\begin{equation}
\frac{\partial n_\alpha(r,t)}{\partial t} = \frac{\partial}{\partial r} D \frac{\partial n_\alpha(r,t) }{\partial r} \approx D(t) \frac{\partial^2 n_\alpha(r,t)}{\partial r^2},
\label{eq:diff}
\end{equation}
with a time-dependent diffusion strength $D(t)$, that is independent of velocity and radius. The two boundary conditions we employ are $\partial n_\alpha(0,t)/\partial r = 0$ and an open boundary at the outer bound of the plasma, $r=a$, allowing outflow of particles. Equation ~(\ref{eq:diff}) is solved numerically with the Crank-Nicolson scheme and applied before the particle distribution is imported into HAGIS. The diffusive process begins at $t=0$ and continues until the initial time point of the wave-particle interaction calculation. For simplicity, the diffusion is assumed to not affect the background plasma, hence damping rates remain unaffected.

The time-dependency in the diffusion coefficient accounts for a continuous healing of the flux surfaces. A recent study on the ASDEX Upgrade tokamak successfully matched experimental data on MMI-injected argon transport with a healing rate at the time-scale of the thermal quench~\cite{linder20AUGtransport}. Therefore we use
\begin{equation}
D(t) =  D_\alpha e^{-t_N}
\label{eq:D}
\end{equation}
with an initial diffusion strength $D_\alpha$. The extrapolation from the study mentioned above assumes the diffusive process to be independent of particle mass and machine size. The initial diffusion strength $D_\alpha$ will be parameterized, covering cases where (1) no significant number of alpha particles are lost to (2) cases where the transport renders the wave-particle interaction increasingly irrelevant. Case (1) is representative of an upper limit for good post-disruption EP confinement in the plasma core~\cite{izzo11reconf}. Such a strongly confining case is particularly of interest, because under such circumstances the (generally faster, core-localized) RE electrons are also not expelled, and bear the risk of generating a dangerous RE beam. A strong enough transport is likely to deconfine the runaway electrons even faster than alphas, and such cases are less of a concern from a mitigation perspective. This assessment will be discussed by the end of the section. Diffusion magnitudes chosen for (2) are deemed realistic from our current knowledge from (medium-sized) tokamak experiments and simulations~\cite{hoelzl21jorek}. Furthermore, we require the alpha particle radial profile to be unaffected by transport prior to the disruption occurring. Effects like electrostatic microturbulence can change the general slowing-down shape into \qmark{bump-on-tail-like} energy distributions~\cite{wilkie2017global,wilkie18micro}, which would also have effects on the wave-particle interaction.

The parameter space $\mathbf{A}$ is set up with
\begin{align*}
D_\alpha  &= [1, 100]~\unit{m^2/s} \\
t_\text{TQ}  &= [1, 3]\unit{ms} \\
n_{e1}  &= [1,1.5,2,2.5,3,3.5,4] n_{e0} \\
n_\text{Ne1} / n_\text{D1}  &= [0, 0.1, 1.0],
\end{align*}

Every parameter in $\mathbf{A}$ affects the particles, meaning we obtain distribution functions $f(v,r,t,\mathbf{A})$ for each of the 84 combinations in the parameter space. The LIGKA-calculated damping remains unaffected by $D_\alpha$ as well as $t_\mathrm{TQ}$, since the initial time-point for the HAGIS simulation remains $t_N=1.5$, i.e a specific temperature profile. Damping rates are shown to be a function of the post-disruption electron density $n_{e1}$ and the neon composition $n_\text{Ne1}/n_\text{D1}$ in figure~\ref{fig:damping2}, while the alpha slowing-down is illustrated in figure~\ref{fig:dis_val}b, where $Z_1$ maintains the role of neon composition.

\begin{figure} [h!]
\centering
\includegraphics[width = 0.95\linewidth]{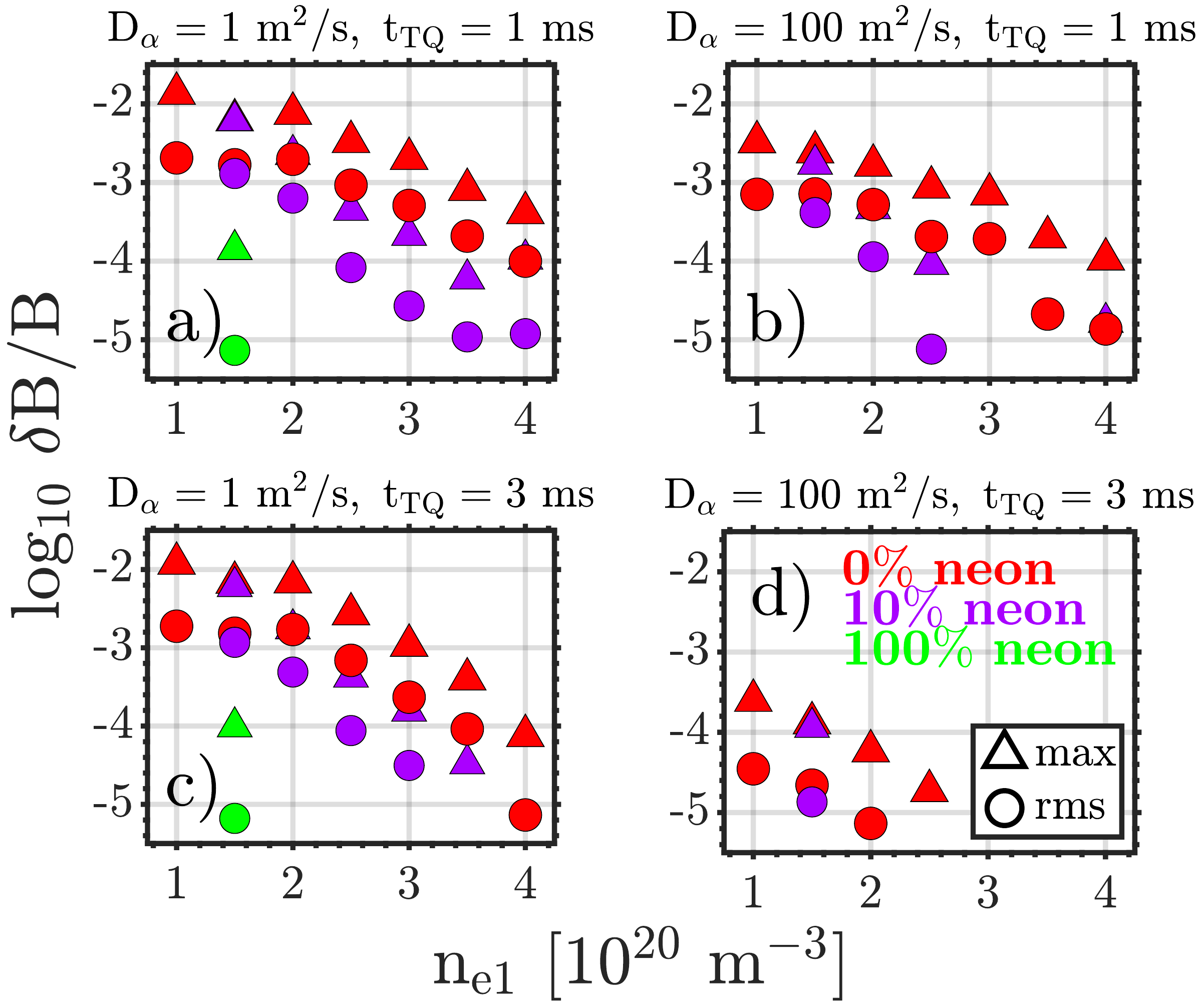}
\caption{Evaluation of maximum (max, triangle) and root-mean-square (rms, circle) of the mode amplitudes $\delta B/B$ reached by post-disruption $M_1$ TAEs resonating with alpha particles. The mean values are calculated after a simulation time of 3~ms. Red, purple and green colors represent 0\%, 10\% and 100\% neon injection, rest deuterium. The simulation is conducted with the HAGIS code where all modes are initialized with an initial perturbation amplitude of $\delta B/B = 10^{-10}$. The initial time-point of the simulation corresponds to a global time of $t_N=1.5$ after the thermal quench has begun. The energetic alpha particle distributions are also calculated for the time-point $t_N=1.5$.}\label{fig:params_a}
\end{figure}

\begin{figure} [h!]
\centering
\includegraphics[width = 0.95\linewidth]{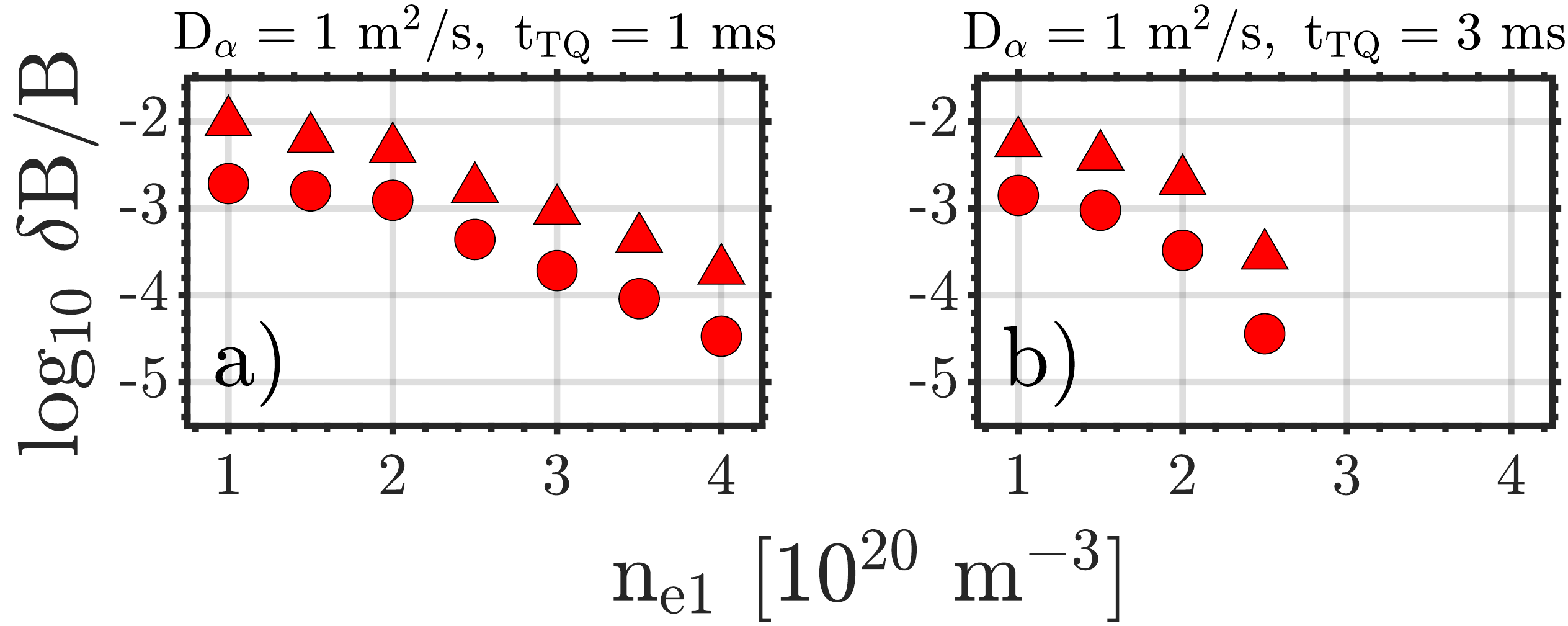}
\caption{Same as figure~\ref{fig:params_a}, but the energetic alpha particle distributions are calculated for the time-point $t_N=3.5$.}
\label{fig:params_b}
\end{figure}

The HAGIS calculations of the wave-particle interactions are conducted with the $M_1$ set of modes and with the same numerical setup described in the previous section. We now evaluate the mode evolution in the parameter space $\mathbf{A}$ in terms of the maximum and the root-mean-square of their amplitudes $\delta B/B$. The main results are collected in figure~\ref{fig:params_a} and figure~\ref{fig:params_b} is part of a sensitivity scan, that will support a discussion on the longevity of the perturbations. In general, we find that alpha particle transport and material injection reduces the perturbation strengths, especially with neon involved.

We begin with a discussion on the left column of figure~\ref{fig:params_a}a and \ref{fig:params_a}c: Most apparent is the general drop in amplitudes with an increase in electron density $n_{e1}$. We have shown that damping rates generally increase with the electron density (figure~\ref{fig:damping2}a)), but have a local minimum at $n_{e1} = 2 n_{e0}$, which in turn creates a local maximum in the parameter space evaluation for $D_\alpha = 1\unit{m^2/s}$. In addition however, a growing electron density also accelerates the slowing-down of alpha particles (figure~\ref{fig:dis_val}b), thereby affecting the drive as well. We can separate the electron density effects on damping and drive by looking at the simulation results for a different thermal quench time $t_\mathrm{TQ}$. The left column of figure~\ref{fig:params_a} shows such a comparison and we see a strong similarity in mode amplitudes for cases, where the electron densities are close to their pre-disruption values. Even though the background temperature profile remains the same, a longer decay time $t_\mathrm{TQ}$ grants the alpha particles more time to decelerate until $t_N = t/t_\mathrm{TQ} = 1.5$ is reached. In the meantime, the damping is independent of $t_\mathrm{TQ}$ as it is a function of the bulk plasma temperature. Hence, the difference we see between the $t_\mathrm{TQ} = 1\unit{ms}$ and $t_\mathrm{TQ} = 3\unit{ms}$ perturbation amplitudes increases with a rising electron density, $n_{e1} \rightarrow 4 n_{e0}$, and is due to the accelerated slowing-down of the alpha particles. 


The amount of neon in the injection modeling is represented by the colors of the markers, red being 0\%, purple 10\% and green 100\% neon. Neon effects on the alpha particles can be captured via the mass-weighted charge $Z_1$ (eq.~(\ref{eq:Z1})). It ranges from $Z_1 = 5/3$ to about $Z_1 \approx 3$ for the largest amounts of neon in $\mathbf{A}$. Figure~\ref{fig:dis_val}b shows, that the effects on the alpha slowing-down are less severe than they are on the damping, in figure~\ref{fig:damping2}. Also, as more neon is added to the plasma, its mass density $\rho$ grows, reducing the Alfv\'en speed $v_A \propto \rho^{-1/2}$ and causing the TAEs to resonate with less energetic particles. All these effects combined cause the observed reduction in perturbation amplitudes, from which we can conclude that the presence of neon is very effective at terminating the alpha-driven TAEs in general.

With an increase in diffusion strength $D_\alpha$ (figures~\ref{fig:params_a}b and \ref{fig:params_a}d), the alpha distribution experiences a stronger flattening of its radial gradient. As the mode growth is driven by the radial pressure gradient we observe the expected drop in overall magnitudes with stronger diffusion. With $D_\alpha = 100\unit{m^2/s}$ and $t_{\rm TQ} = 1\unit{ms}$ the perturbation levels reached are generally an order of magnitude lower than those without diffusion. With an increase in thermal quench time, our real space model assumes a slower healing of the confining flux surfaces, hence a stronger impact of the alpha diffusion. This effect adds to the quench time influence on the alpha particles discussed above and very sufficiently suppresses the TAE mode growth.

The HAGIS model does not incorporate a collision operator for the alpha particles. For a sensitivity study, part of the parameter space simulations are repeated with alpha distributions obtained for a later time-point $t_N=3.5$, i.e. distributions that were longer under the influence of a collision operator in the kinetic calculation that provides the initial distribution for HAGIS. The rest of the simulation setup (esp. damping) remains unchanged, yielding the results shown in figure~\ref{fig:params_b}. Compared to the prior simulations, the alpha particles had an additional 2~ms (for $t_\mathrm{TQ} = 1\unit{ms}$) and 6~ms (for $t_\mathrm{TQ} =3\unit{ms}$) time to be decelerated. Note, that the saturation of mode amplitudes (in the strongest driving, unmitigated case) takes about 1~ms (see figure~\ref{fig:evo}a). In the parameter space regions of good confinement ($D_\alpha \approx 0$), the mode amplitudes are now reduced by up to half an order of magnitude. As one can see from figure~\ref{fig:dis_val}b), the alpha particles at $t_N=3.5$ are close to a complete thermalization and therefore less capable of driving instabilities.

With the parameter space scan we gathered information about the alpha-driven TAEs and how their perturbation amplitudes could be influenced by material injection, alpha particle diffusive transport and varying thermal quench times. The unmitigated and perfectly confining case, $A = (1,1,1,0)$, yields the highest TAE amplitudes. Disruption mitigation systems based on material injection reduce $\delta B/B$ mainly by raising the damping of the bulk plasma. Especially neon is effective at doing so. However, the material injection modeling in this work assumes an instant and uniform deposition at the onset of the TQ. In disruption scenarios where the inner core of the plasma ($s\leq 0.5$) remains close to its pre-disruption condition - hence with a strong alpha particle presence - could therefore expect significant TAE activity during the thermal quench. A non-uniform deposition of material could even enhance the alpha mode drive: a cold front of particles that (relatively) slowly moves inwards or does not penetrate all the way, could potentially raise the alpha spatial pressure gradient and therefore increase the instability drive. While this has not been considered here, the analytical model that has been presented in section~\ref{sec:coll_alpha} would allow an analysis of this effect. 

The wave-particle interaction is sensitive to the transport of alpha particles, which can certainly occur during a disruption-induced break-up of confining flux surfaces. Up to this date, quantifying the post-disruption transport remains an open problem. Some studies suggest~\cite{izzo11reconf} that the stochastic transport during the thermal quench decreases fast with the size of the machine. For ITER, this would mean that the confinement of EPs in the core could be maintained for a significant time after the thermal quench. Core confinement of a RE beam for up to 1~s has been observed at the TCV tokamak~\cite{decker22tcv,tcvreview}, where the RE scenario explicitly relies on the survival of a pre-disruption suprathermal electron seed. We have to note that it is not yet settled whether core confinement is expected in ITER, as some simulations suggest core stochastization leading to runaway loss~\cite{artola22nonaxisymmetric, sarkimaki22confinement}.

The cases of low alpha particle diffusion are representative of the situation of near perfect post-disruption core-confinement. The alpha-TAE interaction discussed in this paper has particular relevance for such cases, as it ultimately comes down to the TAE interaction with the REs (section~\ref{sec:dream}). Due their high speeds runaways, similarly to alpha particles, are susceptible to losses in stochastic magnetic fields. REs have velocities even higher than the alpha ions, therefore it can be assumed that if the disruption-induced breakup of magnetic surfaces is sufficiently strong and sufficiently long, the runaway electron seed losses will be even greater than the alpha particle losses. In cases where the alphas are lost, one may expect that the formation of large RE beams is less likely. However, the RE seed population may be replenished by the constant source of Compton scattering and tritium decay (section~\ref{sec:dream}). 

The following section is dedicated to a self-consistent disruption simulation, where effects of the established TAEs on the generation of runaway electrons is studied. Since the mechanism evolves naturally in the plasma, without need for an external drive, it is an inherent and passive effect. For above reasons, we only consider a scenario, where the healing of the broken up flux surfaces is sufficiently fast to keep both alphas and REs well confined ($D_\alpha = 1\unit{m^2/s}$). Furthermore, we consider the worst-case scenario and look at an unmitigated disruption, which yields the highest perturbation amplitudes.  

\section{Self-consistent simulation of a disrupting plasma generating a RE beam under the influence of alpha-driven TAEs} \label{sec:dream}
\subsection{Calculation of runaway electron transport}
The wave-particle interaction studied in the previous section causes a destabilization of TAEs in the cooling, post-disruption plasma. The perturbation strengths vary widely, but can reach sufficiently high amplitudes that justify further investigation~\cite{lier21alpha,helander00suppression}. This section is dedicated to a study how the alpha-driven TAEs of the unmitigated case, $\mathbf{A} = (1,1,1,0) = (D_\alpha = 1\unit{m^2/s},~t_\text{TQ} = 1\unit{ms},~n_{e1} = n_{e0},~n_\text{Ne1} = 0)$, affect the generation process. We use the larger $M_2$ set of modes (see figure~\ref{fig:evo}).

The first step consists of establishing the particle transport caused by the perturbative TAEs on runaway electrons. A suitable code for the task is ASCOT5~\cite{varje19ascot,scott20ascot}, which performs orbit-following Monte-Carlo simulations on test particles in a perturbed tokamak equilibrium, yielding advection and diffusion coefficients for energetic particles, resolved in radius, energy and pitch. For the second step we will apply DREAM~\cite{hoppe21dream}, a tool that was designed to simulate runaway electron dynamics of a cooling plasma. Crucial for its application here is its ability to include radial particle transport effects into the runaway evolution dynamics. 

The diffusion coefficient is computed in ASCOT5 using test particle tracing~\cite{konsta16ascot}, where markers, that are initially located at the same radial position are traced sufficiently long for their orbits to have become de-correlated ($\approx 0.5\unit{ms}$ in our case). Their time-dependent radial spreading is used to estimate the transport. The diffusion coefficient is evaluated at different radial positions; where at each radial position 500 markers are initialized uniformly in both the toroidal and poloidal angle along the drift surface. The markers represent runaway electrons accelerated by the toroidal electric field and as such are strongly passing with a pitch $\xi = p_\|/p = 0.99$, where $p$ is the relativistic momentum and $p_\|$ is its component alonged the unperturbed magnetic field line (at the minimum value of the magnetic field for a given flux surface). Since the runaway electrons cover a wide range in energy, the simulation is repeated for different electron energies $[0.1,1,10]\unit{MeV}$. The induced electric field itself is not present in the simulation and neither are the Coulomb collisions, so that all the observed transport is due to the magnetic field perturbations.

\begin{figure}
\includegraphics[width = \linewidth]{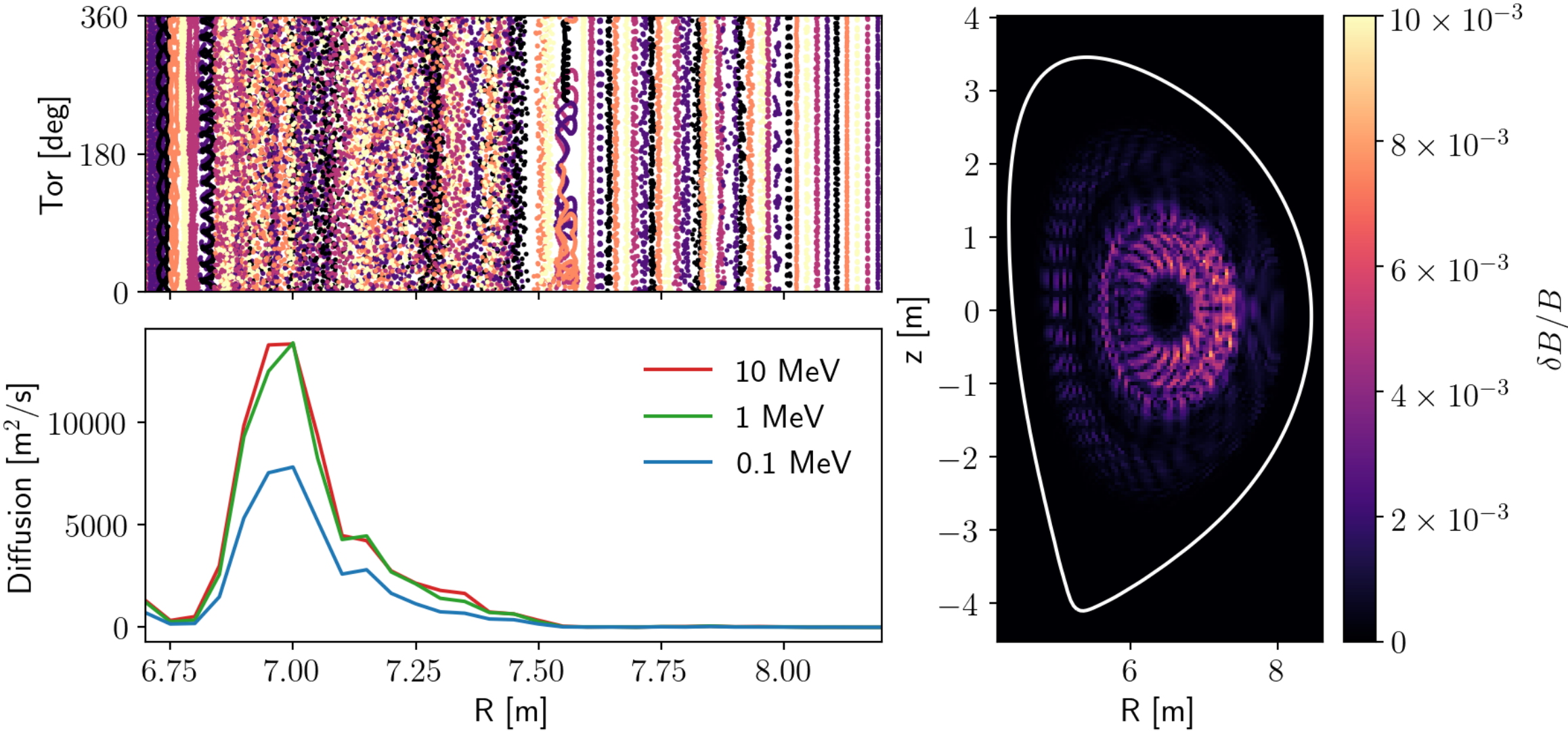}
\caption{Right: the plasma equilibrium perturbed by $M_2$ Alfv\'enic modes (pre-disruption separatrix marked in white). Top left: the Poincare plot for 100~keV electrons, bottom left: the diffusion calculations for test-particles inside these perturbed flux surfaces.}
\label{fig:ascot}
\end{figure}

As we look at a case with $D_\alpha \approx 0$, we can ignore our diffusive model of post-disruption transport, that would otherwise also affect the REs in the plasma. The equilibrium obtained for the LIGKA and HAGIS simulations is converted into a suitable input for ASCOT5, forming a $\kappa = 1.46$ elongated plasma with $R_0=6.2$~m, $a=2.06$~m, surrounded by a conducting wall with the radius $b=3.72$~m.  From the mode evolution of $M_2$ (figure~\ref{fig:evo}b) we calculate the average mode amplitudes RMS$(\delta B/B)_{\mathrm{asc}}$ in the time-frame $t = t_{\mathrm{asc},i} - t_{\mathrm{asc},f}$. At $t_{\mathrm{asct},i}=1.38$~ms the root-mean-square of amplitudes have their maxima and the duration of 0.5~ms is set manually. Eigenfunctions of $M_2$ are set to be constant in time for the duration of the simulation. The perturbed plasma equilibrium is displayed in figure~\ref{fig:ascot}, showing the strongest modes to be localized at the midplane of the plasma, where the radial alpha particle pressure gradient is the strongest. The perturbation amplitude $\delta B/B$ is calculated locally at each grid point, with $\delta B$ being the toroidal maximum of the strength of the 3D MHD-perturbation. The axisymmetric field $B$ and $\delta B$ are both calculated with all their components, i.e. toroidal, poloidal and for the latter also radial contribution. We observe a ballooning-like structure of the perturbations, with a strength $\delta B/B$, that is higher on the low-field side (LFS) of the tokamak. This is both due to the radial dependence of the toroidal magnetic field strength and due to the dominant TAEs being of even parity.

The REs are launched into the perturbed equilibrium and tracked. Figure~\ref{fig:ascot} shows a Poincare plot of 100~keV electrons, with ergodic regions found in $R=6.8- 7.4$~m. With both eigenfunctions and perturbation strengths constant throughout the 0.5~ms simulation, a diffusion strength is calculated as a function of the particle's momentum and radial position, peaking in the ergodic region at $D_{\rm RE} = 13886\unit{m^2/s}$. With first principles~\cite{rr78transport} and the average perturbation strength used ($\delta B/B = 3{\cdot} 10^{-3}$) we can estimate the diffusion a fully stochastic magnetic field would cause at that magnitude, yielding $D_{RR} \approx R_0 c (\delta B/B)^2 \approx 16740 \unit{m^2/s}$. This analytical estimate is known~\cite{konsta16ascot} to overestimate the transport in regions, where magnetic islands occur. However, the transport in this case is caused by ergodization in the \textit{particle phase space} due the the mode overlap~\cite{schneller16alfven}, and not in the underlying magnetic field itself.
We note that there is a remarkable match between the numerical and analytical Rechester-Rosenbluth diffusion result, especially considering the latter uses approximation for the parallel correlation length. Due to the numerical complexity of extracting the numerical diffusion coefficient (for a transport which is approximated by advection-diffusion), it is not unreasonable to assume a 50\% error bar on the numerical diffusion coefficients.

\subsection{Self-consistent modeling of runaway electron dynamics}

In this section we present an analysis with DREAM~\cite{hoppe21dream}, which incorporates the evolution of the background plasma parameters, the induced electric field, and the dynamics of runaway electrons in the presence of radial transport. While DREAM is capable of resolving the runaway electron phase-space, we use it in its ``fluid mode'', where the bulk electrons are characterized by their density $n_e$, temperature $T_e$, and Ohmic current density $j$, while the runaway electrons are characterized by the $j_{\rm RE}$ current density they carry. 

Runaway electron radial diffusion coefficients calculated by ASCOT5 are provided as functions of radius, taken at $\xi=0.99$, and at three normalized relativistic momenta $p \equiv \gamma v/c = \{0.66,\, 2.78,\, 20.5\}$, where $v$ is the electron velocity and $\gamma$ the Lorentz factor. The diffusivities were linearly interpolated within, and extrapolated outside this momentum range: linearly to $0$ at $p=0$, and as constant above the highest provided $p$ point, as well as using a pitch dependence of the diffusivity $D_{\rm RE}\propto |\xi|$, consistently with the Rechester-Rosenbluth diffusion model~\cite{rr78transport}. The obtained $p$ and $\xi$-dependent diffusivities were then translated to a single (only radially varying) diffusion coefficient $D(r)$ for the runaway number density using the method described in Ref.~\cite{svensson20magper}.

The DREAM simulations assume a pure, 1:1 deuterium-tritium plasma, with $n_e$, $T_e$, and $j$ profiles as shown in figure~\ref{fig:pre_dis}. No material injection was done in this simulation. We use a model magnetic equilibrium, equivalent to the Miller parametrization \cite{miller98model}, defined by the major radius at the magnetic axis $R_0$, the plasma minor radius $a$, a wall radius $b$ (representing the closest toroidally closed conducting structural element), and a toroidal field on axis $B_0$. Here we assume the wall to be perfectly conducting, and we set a constant elongation of $\kappa=1.46$, zero triangularity and no Shafranov shift. 

The temperature evolution is prescribed as in eq.~(\ref{eq:T}) with $t_{\rm TQ}=1\,\rm ms$, and a radially constant final temperature of $10\,\rm eV$. The ion charge states are evolved self-consistently accounting for Lyman opacity effects upon recombination \cite{vallhagen2022spi}. The Dreicer runaway generation is calculated using a neural network trained on kinetic simulations~\cite{hesslow19neural}. We also consider primary generation from Compton scattering, tritium decay~\cite{solis17REs,vallhagen20REs}, and hot-tail seed~\cite{svenningsson21ht}. The avalanche growth rate accounts for the partial screening effect~\cite{hesslow2019influence}. The bulk conductivity is calculated using the model by Redl et al.~\cite{redl21conductivity}, that is valid across all collisionality regimes at arbitrary shaping. Note that the runaway generation rates account for magnetic trapping effects, as does the conductivity model with the collisionality dependence of trapping effects accounted for. 



In the simulations we keep the runaway diffusion coefficients constant in time in the entire simulation. Since in reality the magnetic perturbation amplitudes decay in time due to the thermalization of the alphas present, and an increasing damping in the cooling plasma, the results represent an upper bound on the effect of the transport. We also perform a scan over the diffusion strengths, scaling it up and down by $D \times [d_{1000},d_{10},x_1,x_3,x_{10}] = [1/1000,1/10,1,3,10]$, where $x_1$ is the non-scaled, baseline scenario. 

DREAM calculates a runaway rate $\Gamma \equiv \mathrm{d}(n_{\rm RE})/\mathrm{d}t$ for every generation mechanism individually, from which one can obtain the individual runaway current density rates $\mathrm{d} (j_{\rm RE})/\mathrm{d}t = \Gamma e c$. In this equation, every RE travels along the magnetic field lines at the speed of light (RE fluid). The runaway current density rate at each flux surface can be integrated into the runaway current rate $\mathrm{d} I_{\rm RE}/\mathrm{d}t$ of the entire device via
\begin{equation}
\frac{\mathrm{d}I_{\rm RE}}{\mathrm{d}t} = \frac{1}{2 \pi} \int_0^a ~ V' \left\langle \frac{R_0^2}{R^2} \right\rangle \frac{G_{R0}}{B_\mathrm{min}}  \frac{\mathrm{d}j_{\rm RE}}{\mathrm{d}t} \mathrm{d}r, 
\label{eq:I_dream}
\end{equation}
where $\langle \cdot \rangle$ denotes a flux-average value, $R$ is the major radius at any given point, $G_{R0} \equiv (R/R_0) B_\phi$, where $B_\phi$ is the toroidal magnetic field and $B_\mathrm{min}$ is the minimum magnetic field strength on the corresponding flux surface. 

\begin{figure}
\includegraphics[width = 0.48\linewidth]{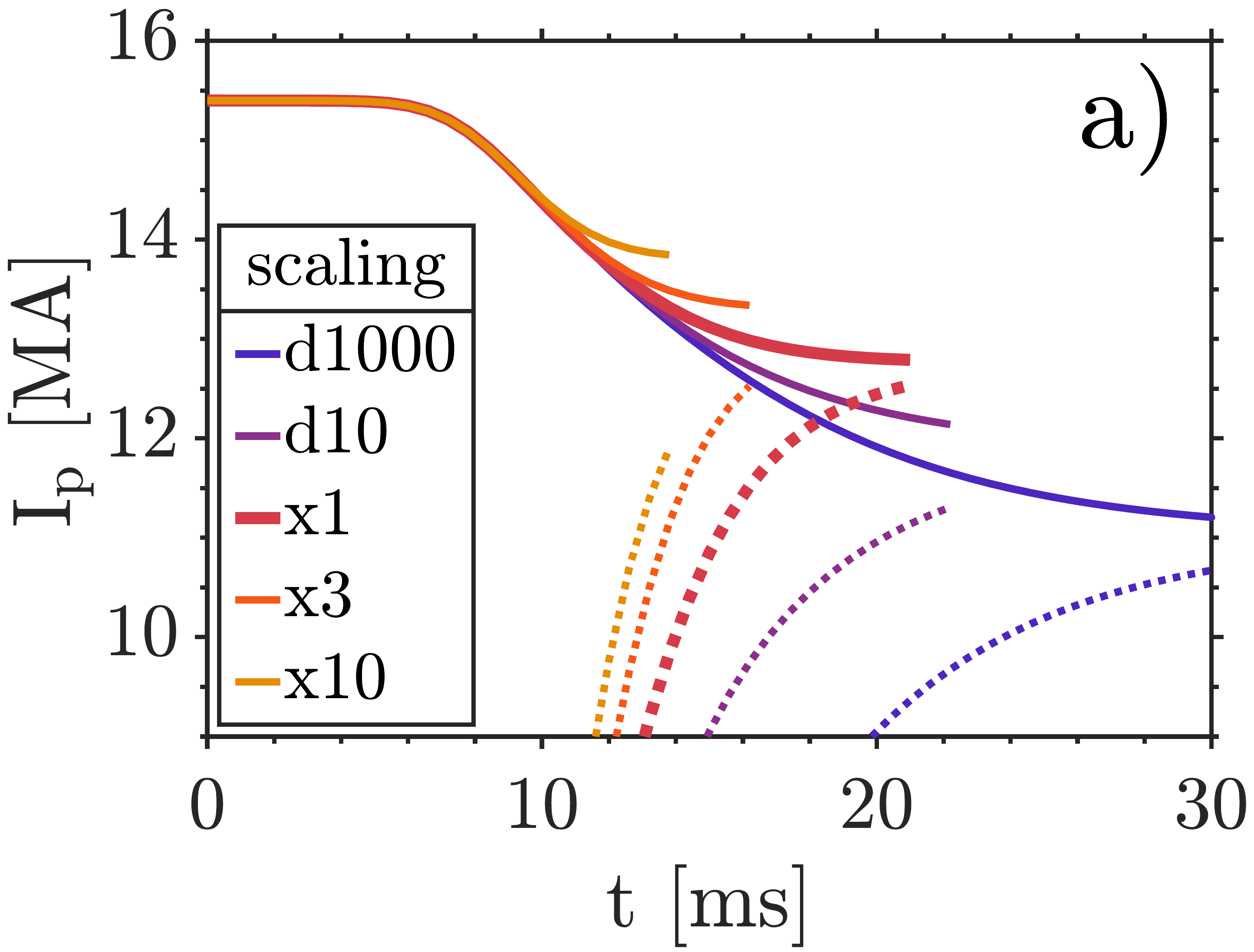}
\includegraphics[width = 0.48\linewidth]{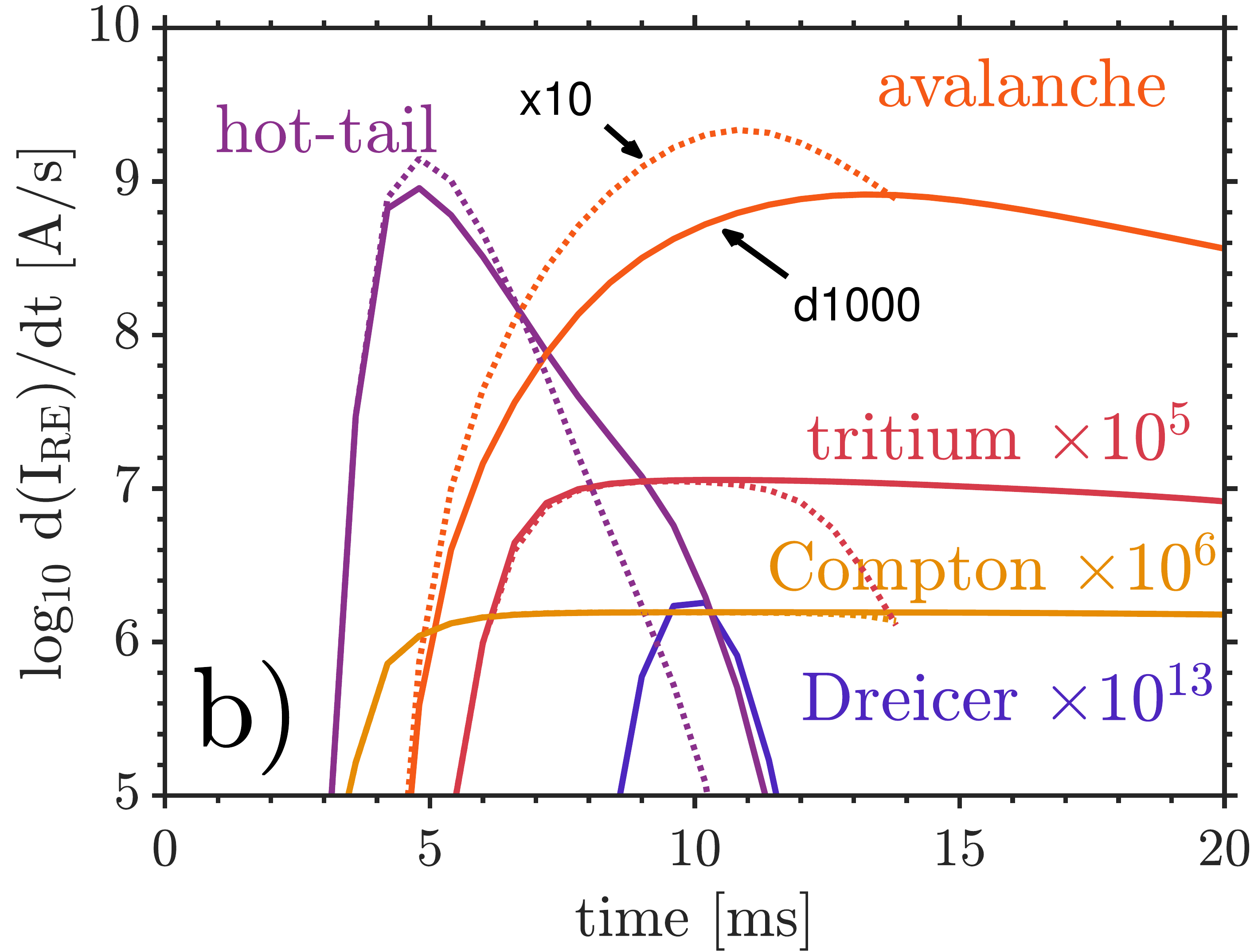} 
\caption{DREAM simulation of an ITER plasma disruption (begins at $t=0$), that is under the influence of alpha-driven TAEs and the RE transport that the TAEs cause. The diffusive transport strength is uniformly scaled up (\labelemph{x}) by factors 3 and 10, and scaled down (\labelemph{d}) by factors of 10 and 1000, while \labelemph{x1} is the baseline scenario. a) Evolution of the total plasma current $I_p$ (solid) and the total runaway current $I_\text{RE}$ (dotted). b) RE generation rates d$(I_\text{RE})/$d$t$ of individual generation mechanisms for up-scaled transport (\labelemph{x10}, dashed) and down-scaled transport (\labelemph{d1000}, solid). Note that the up-scaled simulation does not extend all the way until full conversion.}
\label{fig:dream}
\end{figure} 

Figure~\ref{fig:dream}a shows the evolution of the total plasma current (solid lines) and the total runaway current (dashed) for the nominal transport level (\labelemph {x1}, red), and for the scaled transport coefficients. The simulations show that increasing the strength of the transport (i.e.~moving from \labelemph{d1000} towards \labelemph{x10}) leads to an \emph{increasing} runaway conversion\footnote{Note that the simulation ends before the conversion is complete due to a local sign change in the electric field. This effect cannot be handled in fluid DREAM simulations, and is not changed by increasing resolution.}.

\begin{figure}
\includegraphics[width = 0.48\linewidth]{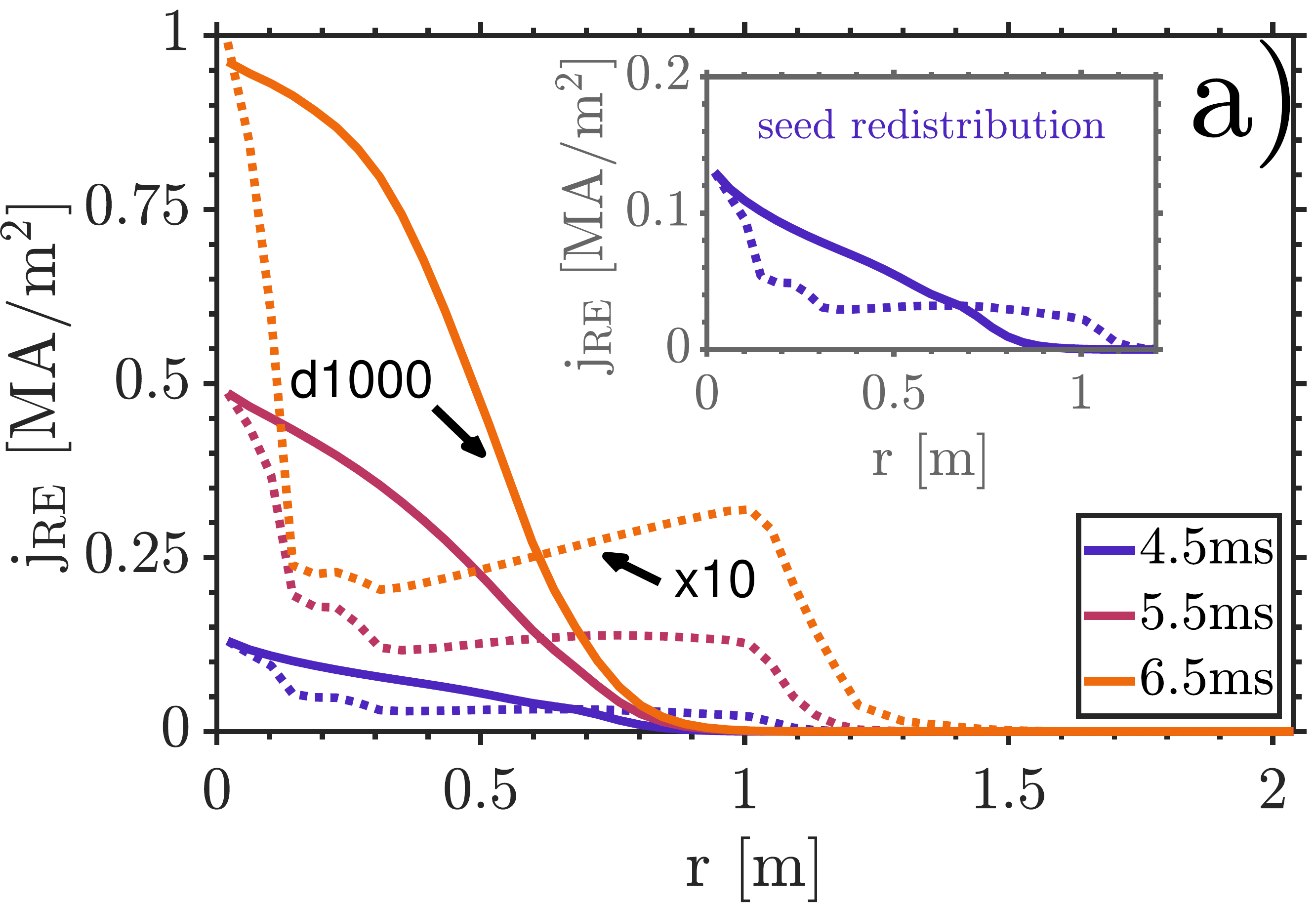}
\includegraphics[width = 0.48\linewidth]{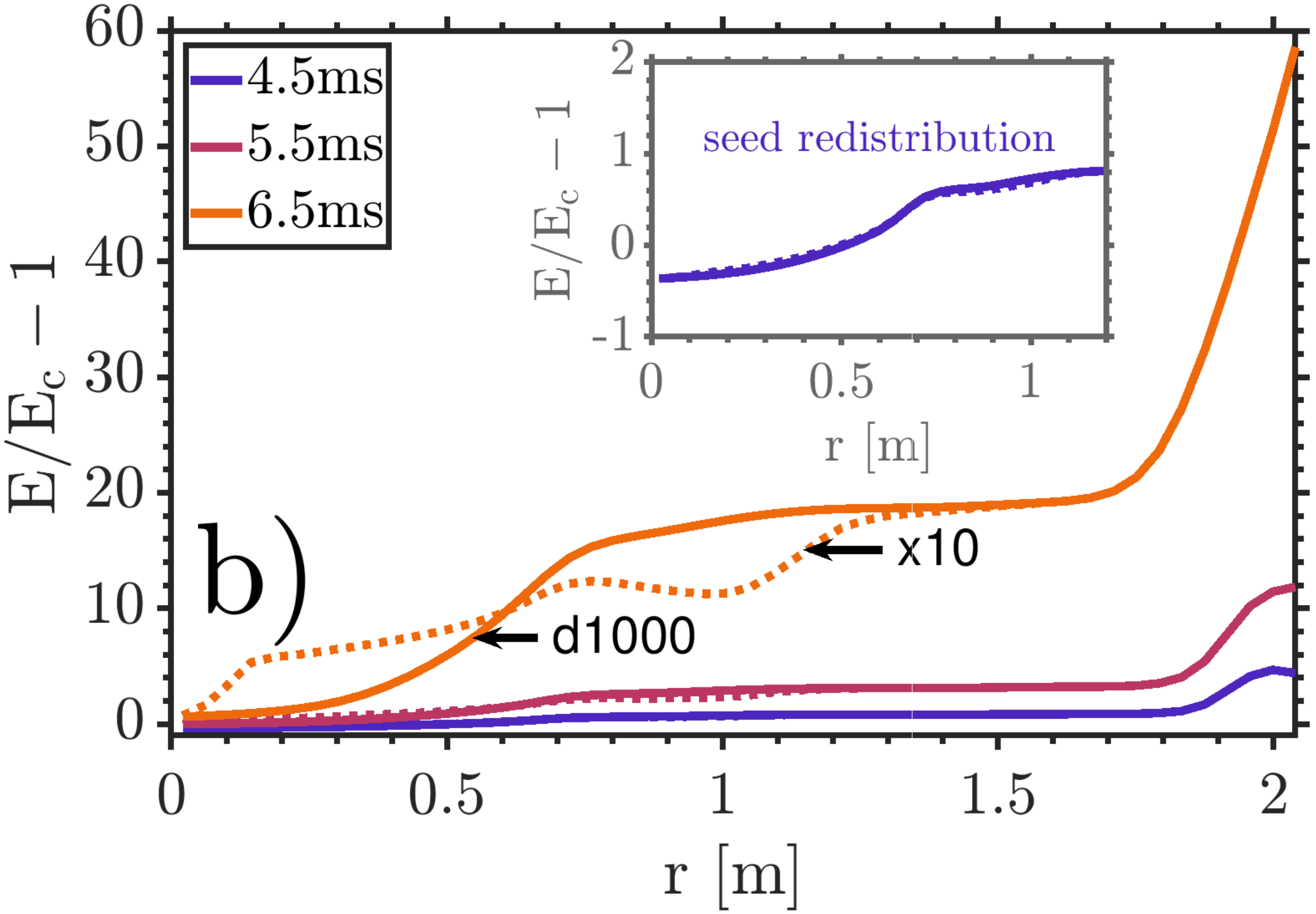}
\includegraphics[width = 0.48\linewidth]{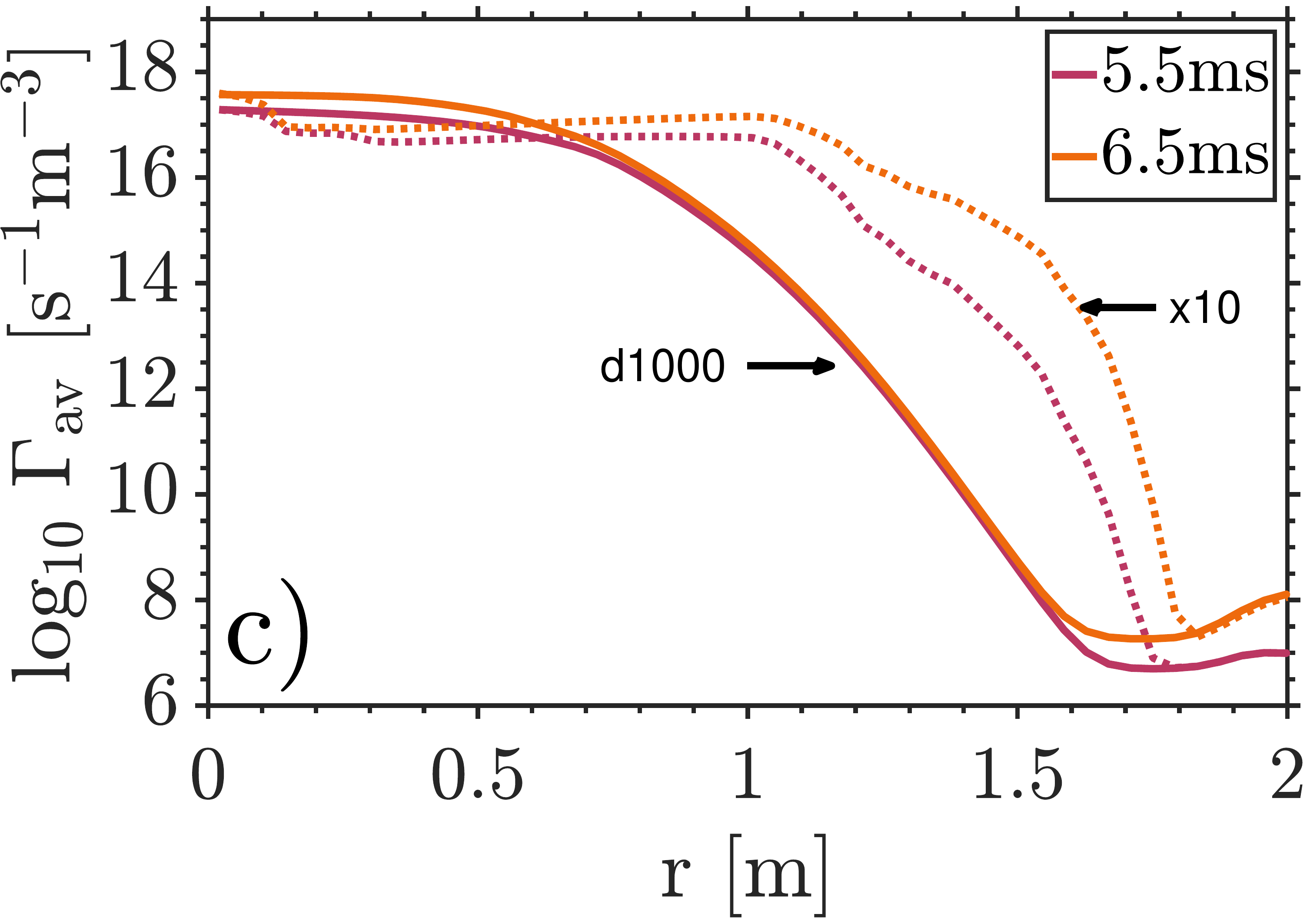}
\caption{DREAM simulation of an ITER plasma disruption (at $t=0$), that is under the influence of alpha-driven TAEs and the RE transport that the TAEs cause. Only the up-scaled RE transport (\labelemph{x10}, dotted) and down-scaled RE transport (\labelemph{d1000}, solid) is shown here for emphasis. Figure a) displays the runaway current density $j_\text{RE}$ b) the the normalized electric field $E/E_c -1$ and c) the avalanche source term $\Gamma_\text{av}$. The time-point $t=4.5\unit{ms}$ approximates the RE seed as it is chosen prior to RE avalanching (see figure~\ref{fig:dream}b) and is additionally zoomed in in a) and b) for emphasis on the seed redistribution effect.}
\label{fig:dream2}
\end{figure}

We take a closer look at the increased transport case (\labelemph{x10}) and compare it to the reduced transport case (\labelemph{d1000}), which is practically unperturbed. Shown in figure~\ref{fig:dream}b are the individual runaway current generation rates d$(I_\text{RE})/\text{d}t$ for every generation mechanism. Note the logarithmic scale on the $y$-axis and the difference in time-scales of the RE current conversion. In both cases, the hot-tail and avalanche are dominating by many orders of magnitude over the other processes. The most significant change caused by the RE transport is found in the avalanche generation. For $t \leq 5\unit{ms}$ the runaway current density $j_\text{RE} \propto \langle n_\text{RE} \rangle$ is a good approximation for the RE seed population, that is eventually multiplied by the avalanche mechanism. In figure~\ref{fig:dream2}a we show how the perturbations redistribute the REs in the TQ. Although the perturbations -- hence diffusion -- do not extend all the way towards $r=0$ (see figure~\ref{fig:ascot}), a significant portion of the core-localized RE seed is transported towards $r \approx 1\unit{m}$, the mid-radius of the plasma. This redistribution has a crucial consequence: as illustrated in figure~\ref{fig:dream2}b the RE seed has been dragged into regions with generally stronger $E/E_c$ normalized electric field, which is a defining factor for the avalanche growth, leading to an increase in the total runaway current. As the simulation progresses, the diffusion is held constant and keeps distributing the REs. In other words, the perturbations redistribute runaway seed from the core to regions where the seed is weak, but the potential for avalanche multiplication is strong. However, the mode-induced perturbations do not extend to the plasma edge, which could increase runaway losses and an eventual drop in RE current.

The simulations also show that the electric field profile undergoes transport-induced changes for $t=6.5\unit{ms}$. The \labelemph{x10} case decreases $E/E_c$ at the mid-radius, but increases it for $r \lesssim 0.5\unit{m}$, because the electric field induction is tied to the changes of the local current. The high-transport case increases $j_\text{RE} (r\approx 1\unit{m})$, hence the local total current decay is decelerated, resulting in a weaker local electric field $E$. As the seed runaways are transported away from the plasma core, the drop in the driving electric field is counteracted by electric field diffusion from the surrounding regions, generating further runaway seed. These effects combined lead to a net increase in runaway electron current.

The transport-induced spatial rearrangement broadens the runaway electron profile and drags REs into regions that are (i) more favourable for avalanching and (ii) would otherwise not be populated by significant number of RE seed electrons. The net of the effect is displayed in figure~\ref{fig:dream2}c via the radial profile of the avalanche source term $\Gamma_\text{av}$. The time-points that are shown not only reflect on a time of significant avalanching, but also lie within the lifetime of the energetic alpha particles (figure~\ref{fig:dis_val}b), which are the reason for the transport in the first place. The RE transport indeed reduces $\Gamma_\text{AV}$ in the plasma core, $r < 0.6\unit{m}$, but this reduction is overcompensated by an increase of $\Gamma_\text{AV}$ for $r > 0.6\unit{m}$, which yields an increased runaway current conversion (figure~\ref{fig:dream}a). In summary, the alpha-driven TAEs may end up increasing the runaway current in the absence of further transport from mid-radius towards the edge.

In the unmitigated and unperturbed ITER disruption, the simulations found (figure~\ref{fig:dream}a, \labelemph{d1000}), that approximately 70\% of the pre-disruption current is eventually converted into runaway current by the end of the current quench. Including the effects of alpha-driven TAEs for the worst-case disruption scenario (unmitigated, well-confining), leads to a current conversion fraction of roughly 85\% and a $12.5\unit{MA}$ RE beam. The disruption simulation was conducted without material injection, which not only provides the strongest perturbation amplitudes, but also isolates the effect of TAEs on the RE generation. Mitigation system effects on the TAEs are shown in figure~\ref{fig:params_a}. With the Rechester-Rosenbluth model~\cite{rr78transport} a reduction of the diffusion by a factor of 1000 (\labelemph{d1000}) can be expected for a reduction of the average perturbation amplitude $\delta B/B$ by a factor of $\sqrt{1000}\approx31.6$.  This is achieved by high amounts of deuterium injection ($n_{e1} > 4 n_{e0}$), alpha particle diffusion of the order of $D_\alpha = 100\unit{m^2/s}$ and slow thermal quenches ($t_\text{TQ} = 3\unit{ms}$) in various combinations. More detailed predictions would require extensive parameter space scans, which are non-trivial. For example: while the alpha particle diffusion $D_\alpha$ reduces the perturbation amplitudes by flattening the alpha pressure gradient, it also causes the alpha particles to reach further towards the edge of the plasma. As a result, the TAE amplitudes may become lower, but modes closer to the edge and perhaps can be more easily destabilized. A transport channel for the REs that extends towards the edge, may cause losses of REs and ultimately reduce the RE current. 

In a similar thought, one could think about exploiting this core-transport with the addition of external efforts. In a recent study~\cite{svenningsson21ht}, mitigated ITER disruptions under the presence of magnetic perturbations were found to have substantial effect on RE dynamics. The simulations in the study assumed resonant magnetic perturbations, which are however only able to effectively penetrate in the edge region ($0.6 < r/a < 1.0$). Depending on the mitigation scenario and perturbation amplitudes, these RMPs -- in combination with various injection schemes -- would sometimes increase the RE current. Conclusively, the study found, that an effective dissipation of the runaway electrons is difficult without additional, significant transport in the center. Within this context, the mechanism investigated in this work could provide synergy effects with externally applied perturbations. 

\section{Summary and discussion\label{sec:discussion}} 
We applied kinetic simulations of the alpha particle distribution function during $15\unit{MA}$ ITER thermal quenches. We find that the thermalization of the suprathermal alphas is delayed by several milliseconds with respect to the bulk temperature drop. While this effect is not sufficient for an alpha runaway, the alphas thermalize after approximately $ t=6 t_{\rm TQ}$ following the disruption, allowing for resonant interaction with TAEs. Simulations including the self-induced electric field show that the alpha particle velocity distribution remains isotropic during the slowing-down process. We found that raising the density of the plasma accelerates the thermalization, with the electron density playing a more significant role that the ion composition. 
At a post-disruption electron density of $10^{21}\unit{m^{-3}}$ the alphas thermalize at $t \approx 4.5 t_{\rm TQ}$ (for $t_\text{TQ} = 1\unit{ms}$ i.e. $4.5\unit{ms}$). Whether this elevated electron density was achieved by pure deuterium injection or with the inclusion of heavier mass ions is found unimportant. Slower thermal quenches leave the alphas more time to decelerate before a certain temperature ($\sim t_N$) is reached. This is important for the alpha-TAE drive because the damping is generally a function of temperature. At quenches as slow as $t_\text{TQ} = 5\unit{ms}$ the alphas thermalize already at $t = 5 t_{\rm TQ}$ ($n_e = 10^{20}\unit{m^{-3}}$) and at $t = 3.25 t_{\rm TQ}$ ($n_e = 10^{21}\unit{m^{-3}}$). 

We calculated post-disruption plasma equilibria and the Alfv\'en modes supported by the system, as well as their damping. The TAEs are calculated to experience strongly decreasing damping as the temperature decays, dropping from levels of $\gamma/\omega \approx 4\%$ down to $0.1/0.2\%$ (for $n=8,9$) only $t_N = 1.5 t_\text{TQ}$ into the thermal quench. Disruption scenarios are considered, where mitigation systems might inject various mixtures of deuterium and neon, whose inclusion generally raises the damping rate, especially for mixtures containing neon. 

The knowledge about alpha particles and TAEs previously gained was joined in section~\ref{sec:wp_int}, whose subject is wave-particle interaction simulations. For the unmitigated case, the alpha particles were shown to resonantly drive the TAEs unstable. The simulations begin at $t_N = 1.5$ due to the low damping rates calculated before. A saturation is reached after an additional $t_N=1.5$, with average perturbation amplitudes reaching $\delta B/B \approx 10^{-3}$. For further wave-particle interaction simulations, a parameter space was created, which covers the effects of density/neon injection, thermal quench time and disruption-induced alpha particle transport. The latter is modelled with a diffusion equation at the onset of the disruption and parameterized with the diffusion coefficient $D_\alpha$. The general observation is, that the unmitigated, well-confining ($D_\alpha \approx 0$) case yields the highest perturbation amplitudes. Addition of material overall decreases the TAE amplitudes, both due to an increase in background damping and due to accelerated alpha slowing-down. The diffusion of alpha particles flattens the spatial gradient, from which the energy in the resonant interaction is drawn, yielding generally lower perturbation strengths (up to an order of magnitude less for $D_\alpha = 100\unit{m^2/s}$, $t_\text{TQ} = 1\unit{ms}$). For slower thermal quenches the alphas are less energetic at $t_N = 1.5$, resulting in slightly lower average $\delta B/B$.

We calculated the impact of the alpha-driven TAEs on RE transport and RE generation. We focused on the unmitigated, well-confining disruption, which is the probably the worst-case scenario from a mitigated perspective. With the particle-following code ASCOT5, the RE transport is calculated. A diffusion strength of up to $D_\text{RE} \approx 14000\unit{m^2/s}$ in the inner half of the plasma was found. 

Finally we calculated the self-consistent runaway electron dynamics in the presence of the TAE-induced RE transport using the code DREAM. The RE diffusion is found to generally increase the final runaway current, with a 15\% increase in the unscaled simulation ($I_\text{RE} \approx 13\unit{MA}$) compared to the down-scaled situation ($D_\text{RE}/1000$, $I_\text{RE} \approx 11\unit{MA}$). The reason is found to be the spatial rearrangement of the runaway electron seed, which is eventually multiplied by the avalanche mechanism. The RE seed is diffused into regions which would otherwise not be populated but are generally more favorable for avalanche. Losses of RE particles are not invoked since the $D_\text{RE}$ is most dominant in the plasma core and does not extend all the way towards the plasma edge. The reason for that is ultimately the central location of the TAE-driving alpha population.

With the ITER disruption research in this work, we have learned about an indirect interaction mechanism between fusion-born alpha particles and runaway electrons, with the mediator being TAEs. While the perturbations were found to increase the runaway electron generation, it bears an interesting opportunity for disruption mitigation systems. Systems like RMPs or passive helical coils~\cite{smith13passive,weisberg21passive} apply externally generated perturbations in order to enhance RE transport. Both systems would benefit from the core-transport mechanisms presented in this study. The next logical step therefore is to include such externally generated perturbations into the disruption simulations alongside the TAEs. In combination with the core-localized TAEs, a synergy effect in reducing the final RE current seems promising. 

Further effects which enhance a fast ion tail may contribute to mode drive - in particular, direct ion heating methods, such as beam heating or ion cyclotron resonance heating. Future analysis will be necessary to quantify the potential impact of external heating on post-disruption runaway dynamics.

\section*{Acknowledgments}

This project has been partially carried out within the EUROfusion ENR ATEP and TSVV-9 projects.
This work has been carried out within the framework of the EUROfusion Consortium, funded by the European Union via the Euratom Research and Training Programme (Grant Agreement No 101052200 -- EUROfusion).  Views and opinions expressed are however those of the author(s) only and do not necessarily reflect those of the European Union or the European Commission. Neither the European Union nor the European Commission can be held responsible for them.
Some of the simulations were performed on the Marconi-Fusion supercomputer hosted at CINECA.

\bibliographystyle{iaea_papp_natbib}
\bibliography{literature}
\end{document}